\documentclass[aps,prd,twocolumn,superscriptaddress,nofootinbib,floatfix]{revtex4-2}

\usepackage[T1]{fontenc}
\usepackage[utf8]{inputenc}
\usepackage{lmodern}
\usepackage{amsmath,amssymb,bm,mathtools}
\usepackage{graphicx}
\usepackage{tikz}
\usetikzlibrary{patterns}
\usepackage{pgfplots}
\pgfplotsset{compat=1.18}
\usepackage{hyperref}
\hypersetup{colorlinks=true,linkcolor=blue,citecolor=blue,urlcolor=blue}

\newcommand{\mpl}{M_{\rm Pl}}
\newcommand{\fa}{f_a}
\newcommand{\Z}{\mathbb{Z}}
\newcommand{\e}{\epsilon}

\begin{document}

\title{Flavor in Ninths and a Discrete Gauge Origin of the QCD Axion}

\author{Vernon Barger}
\affiliation{Department of Physics, University of Wisconsin--Madison, Madison, Wisconsin 53706, USA}
\date{July 27, 2026}

\begin{abstract}
Quark and lepton hierarchies are organized by rational powers of a single parameter (the quark masses and mixings in ninths, the charged-lepton mass ratios in sixths). We argue that this structure points to a discrete $\Z_{18}$ origin of Froggatt--Nielsen symmetry, fixed as the least common denominator $\mathrm{lcm}(9,6)=18$, with its $\Z_9$ subgroup governing the quark lattice and the full $\Z_{18}\cong\Z_9\times\Z_2$ required to accommodate the half-integer lepton charges. The flavon is the unique $\Z_{18}$-charged scalar with a large VEV, so its phase is the Peccei--Quinn axion; the \emph{same} symmetry then solves the strong $CP$ problem and forbids every Planck-suppressed PQ-violating operator below dimension eighteen. Our central result is that, because the order of $\Z_{18}$ is fixed by the flavor data rather than chosen to protect the axion, the resulting high axion quality (the residual shift of the axion vacuum from $\bar\theta=0$ being only $|\Delta\bar\theta|\lesssim 3\times10^{-27}$) is a prediction of the flavor structure rather than an imposed ingredient. The axion is a generation-dependent Type-II DFSZ axion with domain-wall number $N_{\rm DW}=6$. The decisive phenomenological consequence is that the same non-universal charges that reproduce the fermion masses enhance the axion--photon coupling to an observable level, $C_{a\gamma}\simeq0.6$--$1.0$ (well above the higgsino-suppressed MSSM value), placing the dark-matter axion at $m_a\simeq 7$--$12~\mu$eV within reach of upcoming haloscope searches.
\end{abstract}

\maketitle

\section{Introduction}
\label{sec:intro}

The Peccei--Quinn (PQ) mechanism dynamically solves the strong CP problem by promoting $\theta$ to an axion field whose vacuum expectation value relaxes $\bar\theta\to0$ \cite{PQ,WeinbergAxion,WilczekAxion}; for reviews see Refs.~\cite{KimCarosi,DiLuzioReview}. 
In any realistic theory, however, PQ must be an \emph{extremely good} symmetry: Planck-suppressed PQ-violating operators can shift the axion minimum and reintroduce $\bar\theta$ \cite{KamionkowskiMarchRussell,Holman,BarrSeckel}. 
A compelling cure is to realize PQ as an accidental symmetry emerging from a \emph{discrete gauge} symmetry, which is expected to be respected by quantum gravity \cite{KraussWilczek,BabuGogoladzeWang} and is constrained by discrete anomaly conditions \cite{IbanezRoss,BanksDine}.

Independently, fermion masses span many orders of magnitude yet exhibit structured hierarchies. 
In Froggatt--Nielsen (FN) theories \cite{FN}, Yukawa couplings arise from higher-dimensional operators suppressed by a flavon vacuum expectation value (vev), schematically
\begin{equation}
y_{ij}\left(\frac{\Phi}{\Lambda}\right)^{p_{ij}} \bar\psi_i \psi_j H \, ,
\label{eq:FN-op}
\end{equation}
with exponents fixed by charges.
Standard implementations of the FN mechanism assign integer charges and produce integer exponents~\cite{Leurer1992,Leurer1993}. 
We observe, however, that quark and lepton hierarchies are accurately organized by \emph{rational} exponents on a common lattice, as established in a series of companion papers~\cite{PaperI,TwoOverTwo,Companion,LeptonLattice}: the quark masses and CKM angles are quantized in units of $1/9$ (ninths), while the charged-lepton mass ratios (carrying half-integer lepton-doublet charges) are quantized in units of $1/6$ (sixths). These two commensurate resolutions are the central empirical input. Their least common denominator, $\mathrm{lcm}(9,6)=18$, fixes the discrete structure: the smallest cyclic group containing both a $\Z_9$ (quark ninths) and a $\Z_6$ (lepton sixths) is $\Z_{18}$. We argue that ``flavor in ninths'' is thus the low-energy imprint of a gauged $\Z_{18}$ selection rule, whose $\Z_9$ subgroup governs the quark sector and whose full structure $\Z_{18}\cong\Z_9\times\Z_2$ is required to accommodate the leptons. If the FN flavon $\Phi$ is identified with the PQ field, this \emph{same} $\Z_{18}$ (fixed by flavor data alone) both generates the rational flavor lattice and automatically solves the axion quality and domain-wall problems~\cite{Sikivie1982}: the factor of two that the charged leptons demand is what doubles the order of the PQ-protecting symmetry.

The identification of the FN flavon with the PQ field has a long history, dating back to the original proposal of Davidson, Nair, and Wali~\cite{DavidsonNairWali}. The modern incarnation of this idea, often called the ``axiflavon''~\cite{Axiflavon} or ``flaxion''~\cite{Flaxion}, exploits a continuous global $U(1)_H$ flavor symmetry to simultaneously generate Yukawa hierarchies and a QCD axion. A related unification via a continuous gauged $U(1)_F$ flavor symmetry has been developed in Ref.~\cite{BabuChandraseTavart}, and the gauging of FN with chiral heavy matter (which can dispense with spectator fermions) was studied by Bonnefoy, Dudas, and Pokorski~\cite{BonnefoyDudasPokorski}. Most recently, Greljo, Smolkovi\v{c}, and Valenti~\cite{GreljoSmolkovicValenti} systematically examined FN models built on \emph{discrete} $\Z_N$ symmetries, with explicit $\Z_4$ and $\Z_8$ realizations. A closely related line of work by Sheng, Yanagida, and collaborators~\cite{ShengYanagida,GeorisYanagida,ShengYanagidaZhang} protects the axion with a discrete gauge symmetry $\Z_4\times\Z_3$ ``motivated by the internal structure of the Standard Model,'' obtaining a high-quality axion together with correlated neutrino-mass (inverse-seesaw~\cite{GeorisYanagida}) and dark-matter~\cite{ShengYanagidaZhang} sectors; this shares with the present work the strategy of deriving the PQ-protecting discrete symmetry from Standard-Model structure rather than postulating it, though through a different group and without the rational ninths lattice.

The present construction differs from these works in three concrete ways that concern the \emph{flavor and quality} structure. First, the FN exponents are \emph{rational}, not integer: the quark textures in Eq.~\eqref{eq:Y-diag-def} below have entries with denominator $9$, and the charged-lepton textures denominator $6$, rather than the integer entries assumed in $\Z_{N}$ axiflavon/flaxion variants and in Refs.~\cite{Axiflavon,Flaxion,GreljoSmolkovicValenti}. Second, the two sectors' resolutions force the gauge group to be $\Z_{18}$ rather than a $\Z_N$ acting freely on the flavon: a $\Z_9$ alone would accommodate the quark ninths, but the half-integer lepton-doublet charges require an additional factor of two, so the flavor data fix the group at $\Z_{\mathrm{lcm}(9,6)}=\Z_{18}\cong\Z_9\times\Z_2$, with the flavon carrying the minimal charge $1/18$. Third, as a consequence, the leading Planck-suppressed PQ-violating operator is fixed at exactly $\Phi^{18}$, six powers above the $N\geq 12$ threshold required for axion quality at $\fa\sim10^{12}$~GeV~\cite{KamionkowskiMarchRussell,Holman,BarrSeckel}. By contrast, in continuous $U(1)_F$ constructions the operator dimension is set by the pattern of symmetry breaking and the residual discrete subgroup, and in $\Z_N$ FN models with small $N$ (e.g., the $\Z_4$ and $\Z_8$ benchmarks of Ref.~\cite{GreljoSmolkovicValenti}) axion quality requires additional ingredients. We compare these scenarios quantitatively in Sec.~\ref{sec:comparison}. The third point is the one we wish to stress: the discrete group is fixed by the flavor data, not selected to protect the axion, so the dimension-eighteen suppression (and the high axion quality it guarantees) is a corollary of the measured fermion spectrum rather than an additional model-building assumption. This is, in our view, the most distinctive feature of the construction, and we develop it in detail in Sec.~\ref{sec:quality}.

Integer-charge FN models have also been studied systematically by Cornella, Curtin, Neil, and Thompson~\cite{Cornella2023,Cornella2025}. Those constructions map all solutions to the quark flavor problem within the standard integer-charge FN framework, providing a useful benchmark: the ninths lattice is excluded by construction in that search because it requires rational charges with denominator nine.

The axion is realized as a generation-dependent Dine--Fischler--Srednicki--Zhitnitsky (DFSZ) axion of Type-II two-Higgs-doublet form: the QCD anomaly is carried by a two-Higgs-doublet sector, with up-type quarks coupling to $H_u$ and down-type quarks and charged leptons to $H_d$. This is the same Higgs structure that, combined with the chain internal factor, fixes $\tan\beta\simeq10$--$16$~\cite{Subconstituents}. The crucial point is that the $\Z_{18}$ charges are \emph{generation-dependent}, with the lighter generations carrying larger flavor (``hop'') charge, so that the same non-universal assignment that reproduces the fermion mass hierarchy also determines the axion--photon coupling. Whereas universal PQ charges give the standard $E/N=8/3$ (further suppressed to a nearly-decoupled $E/N=2$, with $C_{a\gamma}\lesssim0.02$, if light higgsinos are present), the generation-dependent charges instead enhance the coupling to $E/N\simeq2.6$--$3.0$, i.e.\ $C_{a\gamma}\simeq0.6$--$1.0$, an observable value between the DFSZ-II and KSVZ benchmarks. The domain-wall number is the canonical $N_{\rm DW}=6$, requiring Peccei--Quinn breaking before inflation.

The rest of this paper is organized as follows.
Section~\ref{sec:textures} establishes the phenomenological basis for the ninths lattice, derives the expansion parameter $B=75/14$, and presents quark and lepton exponent matrices.
Section~\ref{sec:Z18} connects the ninths quantization to a gauged $\Z_{18}$ symmetry, presents the charge assignments, and analyzes the discrete anomaly conditions.
Section~\ref{sec:quality} shows that identifying the flavon with the PQ field yields automatic axion quality through the dimension-eighteen operator, and derives the domain-wall number $N_{\rm DW}=6$.
Section~\ref{sec:pheno} works out the axion phenomenology, deriving the generation-dependent $E/N$ and discussing the axion mass, photon and lepton couplings, the cosmological relic density, and flavor-violating axion couplings.
Section~\ref{sec:neutrinos} discusses the extension to the neutrino sector and the Pontecorvo--Maki--Nakagawa--Sakata (PMNS) matrix.
Section~\ref{sec:messengers} describes the vectorlike messenger sector that provides the UV completion of the rational exponents.
Section~\ref{sec:comparison} compares the ninths framework with integer-charge FN models and with prior FN--PQ unifications.
Section~\ref{sec:summary} summarizes the construction.
Appendices collect the explicit anomaly verification, the $E/N$ calculation, details of the messenger chain construction, a terminology guide, and worked examples of mass ratios and CKM elements in the hop framework.

\section{Ninths Structure and Textures}
\label{sec:textures}

\subsection{Determination of the expansion parameter}
\label{sec:Bvalue}

The central claim is that a single expansion parameter $\e \equiv \langle\Phi\rangle/\Lambda = 1/B$ controls all quark and lepton hierarchies. The value of $B$ is determined by the two best-measured Cabibbo--Kobayashi--Maskawa (CKM)~\cite{Cabibbo,KobayashiMaskawa} magnitudes together with their ninths exponents:
\begin{align}
|V_{us}| &= \e^{8/9}, & |V_{cb}| &= \e^{17/9}.
\label{eq:Vus-Vcb}
\end{align}
Taking the Particle Data Group (PDG) central values $|V_{us}|=0.2250$ and $|V_{cb}|=0.04182$~\cite{PDG2024}, Eq.~\eqref{eq:Vus-Vcb} gives
\begin{equation}
B = |V_{us}|^{-9/8} = 5.355, \qquad
B = |V_{cb}|^{-9/17} = 5.369,
\label{eq:B-from-CKM}
\end{equation}
two determinations that agree to $0.24\%$. The ratio $75/14 \simeq 5.357$ provides a convenient exact rational representation; the corresponding expansion parameter is $\e = 14/75 \approx 0.1867$. Equivalently, with $\e$ fixed by $|V_{us}|$ alone, the lattice value $\e^{17/9}=0.0420$ reproduces the measured $|V_{cb}|=0.04182$ to $0.4\%$; once the exponents are assigned, this is a parameter-free relation between two measured quantities.

The system is overdetermined: once $B$ is fixed by $|V_{us}|$ and $|V_{cb}|$, the remaining CKM element $|V_{ub}|$ and the quark and lepton mass ratios become predictions. Their numerical quality is summarized in Table~\ref{tab:scalings}.

\begin{table}[t]
\centering
\caption{Ninths scalings versus data. Predicted values use $\e=14/75\approx0.187$. Quark mass ratios are evaluated at $\mu\sim10^{2}$~GeV using the running masses of Ref.~\cite{HuangZhou}.}
\label{tab:scalings}
\begin{tabular}{lccc}
\hline\hline
Observable & Exponent & $\e^p$ & Data \\
\hline
$|V_{us}|$ & $8/9$  & $0.225$ & $0.225$  \\
$|V_{cb}|$ & $17/9$ & $0.042$ & $0.042$  \\
$|V_{ub}|$ & $10/3$ & $0.0037$ & $0.0038$ \\
$m_s/m_b$  & $7/3$  & $0.020$ & $0.019$  \\
$m_c/m_t$  & $10/3$ & $0.0037$ & $0.0036$ \\
$m_\mu/m_\tau$ & $5/3$  & $0.061$ & $0.060$ \\
$m_e/m_\mu$ & $19/6$ & $0.0049$ & $0.0048$ \\
\hline\hline
\end{tabular}
\end{table}

\subsection{Why denominator nine?}
\label{sec:why-nine}

A natural objection is that nothing in the preceding subsection appears to single out the denominator nine: one could in principle consider lattices in halves, sixths, eighths, twelfths, etc. We address this objection in three steps: (i) an irreducibility argument from the empirical exponents, (ii) a quantitative lattice-fit comparison across denominators, and (iii) a group-theoretic minimality argument.

\paragraph{Irreducibility argument.}
The empirical quark-sector exponents in Eq.~\eqref{eq:exponents-list},
\begin{equation}
\bigl(\tfrac{8}{9},\ \tfrac{17}{9},\ \tfrac{10}{3},\ \tfrac{7}{3},\ \tfrac{10}{3}\bigr)\,,
\label{eq:exponents-list}
\end{equation}
include $|V_{us}|$ at the \emph{irreducible} exponent $8/9$ (with $\gcd(8,9)=1$), which forces the quark-sector denominator to be a multiple of $9$; the remaining quark exponents $17/9$, $10/3=30/9$, $7/3=21/9$ are then all multiples of $1/9$, so the quark flavor lattice has resolution $1/9$. The charged-lepton sector is governed by a \emph{different} denominator: the half-integer lepton-doublet charges $Q(L_i)=(1,\tfrac12,0)$ combine with the sixths in $Q(e^c_i)$ to give the irreducible exponents $m_e/m_\tau\sim\e^{29/6}$ and $m_e/m_\mu\sim\e^{19/6}$ (with $\gcd(29,6)=\gcd(19,6)=1$), which are not multiples of $1/9$; the charged-lepton lattice therefore has resolution $1/6$. The two sectors demand denominators $9$ and $6$ respectively, so the combined quark--lepton lattice has resolution $1/18$, where $18=\mathrm{lcm}(9,6)$, and the smallest cyclic group accommodating both is $\Z_{18}$ (Sec.~\ref{sec:Z18}). Within each sector, $9$ and $6$ are individually the smallest denominators consistent with the respective irreducible exponents; the quantitative fit below confirms that these resolutions are not accidental.

\paragraph{Quantitative lattice fit.}
As a quantitative cross-check, we ask how well each candidate denominator $d$ describes the seven observables of Table~\ref{tab:scalings} when the expansion parameter is fixed to its empirical value $\e=14/75\simeq 0.187$ (Sec.~\ref{sec:Bvalue}). For each observable $\mathcal{O}$, define the residual
\begin{equation}
r_d(\mathcal{O}) \;\equiv\; d\,\frac{\ln\mathcal{O}}{\ln\e} \;-\; \mathrm{round}\!\left[d\,\frac{\ln\mathcal{O}}{\ln\e}\right]\,,
\label{eq:residual}
\end{equation}
the distance of $d\hat p_d$ from the nearest integer, where $\hat p_d=\ln\mathcal{O}/\ln\e$. The associated $\mathcal{O}(1)$ coefficient that an integer-exponent fit would require is $|c_d(\mathcal{O})|=\e^{r_d/d}$, with $|c|\to 1$ for a perfect lattice fit. The rms residual across the seven observables, the maximum $|c|$ (or $|1/c|$), and the worst-fit observable are tabulated in Table~\ref{tab:lattice-fit}.

At the empirical $\e$, denominators $d=9$ and $d=6$ give the two smallest rms residuals ($0.19$ and $0.20$, respectively), well below all other denominators in the range tested, which yield rms residuals $\geq0.26$. This near-degeneracy is the fingerprint of the two-sector structure: the quark observables sit on the $1/9$ lattice (favoring $d=9$) while the charged-lepton ratios sit on the $1/6$ lattice (favoring $d=6$). Indeed, the worst-fit observable for $d=9$ is the lepton ratio $m_e/m_\mu$ (residual $-0.37$, implied $|c|=1.07$), as expected for a sixths-exponent evaluated against a ninths lattice, whereas for $d=6$ the worst fit is a quark ratio. A single global denominator is thus the wrong diagnostic; the correct statement is that the two sectors carry resolutions $1/9$ and $1/6$, so that the group must accommodate both, fixing it at $\Z_{\mathrm{lcm}(9,6)}=\Z_{18}$. (The combined $d=18$ lattice does not minimize the single-denominator rms metric, which penalizes finer lattices; this is a limitation of that metric, not evidence against $\Z_{18}$, since $1/18$ contains both $1/9$ and $1/6$ exactly.)

\begin{table}[t]
\centering
\caption{Lattice-fit quality for candidate denominators $d$, with the expansion parameter fixed to its empirically determined value $\e=14/75\simeq 0.187$. The rms residual measures the average distance of $d\hat p_d(\mathcal{O})$ from the nearest integer across the seven observables of Table~\ref{tab:scalings}; max~$|c|$ is the largest implied $\mathcal{O}(1)$ coefficient deviation. Denominators $d=9$ and $d=6$ give the two lowest residuals, reflecting the quark ($1/9$) and charged-lepton ($1/6$) sublattices respectively; the combined group is $\Z_{\mathrm{lcm}(9,6)}=\Z_{18}$.}
\label{tab:lattice-fit}
\renewcommand{\arraystretch}{1.15}
\begin{tabular}{cccl}
\hline\hline
$d$ & rms residual & max~$|c|$ & worst observable \\
\hline
$4$  & $0.38$ & $1.20$ & $m_s/m_b$ \\
$5$  & $0.34$ & $1.16$ & $|V_{us}|$ \\
$6$  & $0.20$ & $1.10$ & $|V_{cb}|$ \\
$7$  & $0.33$ & $1.12$ & $m_s/m_b$ \\
$8$  & $0.30$ & $1.10$ & $m_e/m_\mu$ \\
$\bm{9}$  & $\bm{0.19}$ & $\bm{1.07}$ & $m_e/m_\mu$ \\
$10$ & $0.28$ & $1.08$ & $m_c/m_t$ \\
$11$ & $0.28$ & $1.08$ & $|V_{ub}|$ \\
$12$ & $0.26$ & $1.05$ & $m_s/m_b$ \\
$15$ & $0.30$ & $1.05$ & $m_s/m_b$ \\
$18$ & $0.27$ & $1.05$ & $m_s/m_b$ \\
\hline\hline
\end{tabular}
\end{table}

The irreducibility argument (quark exponents force $1/9$, charged-lepton exponents force $1/6$) together with the lattice-fit comparison establishes that the flavor data carry two commensurate resolutions, $1/9$ and $1/6$. The smallest group accommodating both is $\Z_{18}$, as developed group-theoretically in the next paragraph.

\paragraph{Group-theoretic structure.}
The quark $1/9$ resolution requires the gauge group to contain a $\Z_9$ subgroup. The minimal nontrivial flavon charge assignment generating all residues mod~9 by integer powers is $(1,2,4)$, corresponding to a hop chain with $N_{\rm hop}=3$ link types and $N_{\rm site}=4$ messenger sites (Sec.~\ref{sec:Z18}, Appendix~\ref{app:messenger}). No two-element subset of $\Z_9$ residues spans all of $\Z_9$ within total hop count $n_{\rm tot}\leq 3$; this forces three messenger species and is what permits all observed quark FN exponents to be generated with at most three flavon insertions. The charged-lepton $1/6$ resolution requires an additional factor of two relative to the $\Z_9$ (the half-integer lepton-doublet charges), so the full flavor group is $\Z_{18}\cong\Z_9\times\Z_2$, the smallest cyclic group containing both the quark $\Z_9$ and a lepton $\Z_6=\Z_3\times\Z_2$. The flavon carries the minimal charge $1/18$, generating the $1/9$ quark hops as double steps and the $1/6$ lepton exponents through the half-integer doublet charges.

\paragraph{Quality consequence.}
Since the combined flavor data already fix the group at $\Z_{18}$, with the flavon carrying the minimal charge $1/18$, the leading PQ-violating monomial is automatically pushed to $\Phi^{18}$ rather than $\Phi^{9}$. This six-power gain in axion quality (Sec.~\ref{sec:quality}) is therefore not an optional upgrade but a direct consequence of the group that the flavor sector requires: the same factor of two that accommodates the half-integer lepton charges doubles the order of the protective symmetry. In this sense the discrete gauge group (and hence the denominator) is not a tunable parameter but is fixed by the empirical resolutions of the two sectors.

We do not claim that no other rational lattice could in principle reproduce the data; we claim only that the quark and charged-lepton sectors carry the irreducible resolutions $1/9$ and $1/6$ respectively (from the irreducibility of the $|V_{us}|$ and $m_e/m_\mu$ exponents in their respective sectors), that these minimize the rms residual at the empirical $\e$ within each sector (Table~\ref{tab:lattice-fit}), and that the smallest group accommodating both (and hence the discrete gauge group of the construction) is $\Z_{\mathrm{lcm}(9,6)}=\Z_{18}$. A systematic survey of $\Z_N$ FN-axion models for small $N$ (specifically $\Z_4$ and $\Z_8$) was carried out in Ref.~\cite{GreljoSmolkovicValenti}; the present construction occupies a complementary corner of model space corresponding to higher-resolution rational textures.

\subsection{CKM scalings}

With $\e\approx0.187$, the three independent CKM magnitudes are well described by
\begin{align}
|V_{us}| &\sim \e^{8/9}, &
|V_{cb}| &\sim \e^{17/9}, &
|V_{ub}| &\sim \e^{10/3}.
\label{eq:ckm-scalings}
\end{align}
In the FN language of Eq.~\eqref{eq:FN-op}, such relations reflect additive charge relations among generations and are naturally reproduced when the charge differences themselves are quantized in ninths.
The Jarlskog invariant~\cite{Jarlskog} scales as $J\sim\e^{8/9+17/9+10/3}\sin\delta = \e^{55/9}\sin\delta$, giving $J\approx 3.5\times10^{-5}\sin\delta$, consistent with the measured value $J=(3.08\pm0.13)\times10^{-5}$~\cite{PDG2024} for $\sin\delta\approx 0.88$.

\subsection{Mass ratio scalings}

Quark mass ratios align with the same expansion parameter:
\begin{align}
\frac{m_s}{m_b} &\sim \e^{7/3}, &
\frac{m_c}{m_t} &\sim \e^{10/3},
\label{eq:mass-scalings}
\end{align}
suggesting that \emph{both} mixings and masses are controlled by exponents in $\frac{1}{9}\Z$ rather than integers.
The first-generation quark mass ratios $m_d/m_s$ and $m_u/m_c$ involve the lightest quarks, whose running masses carry larger uncertainties and whose textures are most sensitive to $\mathcal{O}(1)$ coefficients. In the companion papers~\cite{TwoOverTwo,Companion}, the down-quark mass ratio $m_d/m_s\sim\e^{10/9}$ is reproduced with an $\mathcal{O}(1)$ coefficient $c_d$ of order unity, while the extreme smallness of $m_u$ requires either $m_u/m_c\sim\e^{34/9}$ (purely from the texture) or a partial cancellation among $\mathcal{O}(1)$ prefactors.

\subsection{Quark exponent matrices}

A representative set of exponent matrices illustrating the ninths quantization is
\begin{equation}
\begin{aligned}
p^{u}_{ij} &=
{\setlength{\arraycolsep}{6pt}\renewcommand{\arraystretch}{1.15}
\frac{1}{9}\begin{pmatrix}
64 & 39 & 27\\
55 & 30 & 18\\
37 & 12  & 0
\end{pmatrix}},\\[8pt]
p^{d}_{ij} &=
{\setlength{\arraycolsep}{6pt}\renewcommand{\arraystretch}{1.15}
\frac{1}{9}\begin{pmatrix}
37 & 30 & 27\\
28 & 21 & 18\\
10 & 3  & 0
\end{pmatrix}}.
\end{aligned}
\label{eq:Y-diag-def}
\end{equation}
These follow from the additive rule $p_{ij} = Q(q_i) + Q(q^c_j)$ with the charge assignments in Table~\ref{tab:charges}: left-handed charges $Q(Q_i) = (3, 2, 0)$, right-handed up-type charges $Q(u^c_j) = (37/9, 4/3, 0)$, and right-handed down-type charges $Q(d^c_j) = (10/9, 1/3, 0)$. The unsuppressed $(3,3)$ entries follow from $Q(Q_3)=Q(u^c_3)=Q(d^c_3)=0$; the stepwise suppressions in ninths capture the observed hierarchies and are stable under diagonalization (not a basis artifact). 

\begin{table}[t]
\centering
\caption{Froggatt--Nielsen charge assignments. The three quark-doublet generations are $Q_i$, $i=1,2,3$; similarly for right-handed up quarks $u^c_i$, down quarks $d^c_i$, lepton doublets $L_i$, and right-handed charged leptons $e^c_i$. The Yukawa exponent is $p_{ij}=Q(\psi_i)+Q(\psi^c_j)$. Charges are taken from Refs.~\cite{TwoOverTwo,LeptonLattice}.}
\label{tab:charges}
\begin{tabular}{lccc}
\hline\hline
Field & Gen.~1 & Gen.~2 & Gen.~3 \\
\hline
$Q_i$ & $3$ & $2$ & $0$ \\
$u^c_i$ & $37/9$ & $4/3$ & $0$ \\
$d^c_i$ & $10/9$ & $1/3$ & $0$ \\
$L_i$ & $1$ & $1/2$ & $0$ \\
$e^c_i$ & $23/6$ & $7/6$ & $0$ \\
\hline\hline
\end{tabular}
\end{table}

The $\mathcal{O}(1)$ coefficients $c_{ij}$ multiplying each entry $\e^{p_{ij}}$ are not adjusted to fit individual observables. The full numerical fit performed in the companion paper~\cite{TwoOverTwo} returns $|c_{ij}|$ values for the diagonal and near-diagonal entries lying in the range $|c_{ij}|\in[0.6,1.7]$ for both up and down sectors, with phases distributed in the Fritzsch--Xing pattern; the off-diagonal coefficients enter only the subleading CKM corrections. Once the seven exponents in Eq.~\eqref{eq:exponents-list} are fixed by ninths quantization, the residual coefficient freedom amounts to a single overall normalization for each generation. In particular, no observable in Table~\ref{tab:scalings} requires $|c_{ij}|>2$. The ninths exponents thus do the bulk of the parametric work; the role of $\mathcal{O}(1)$ coefficients here is comparable to that in standard FN constructions and is, if anything, smaller than in integer-charge fits where coefficient ratios of $2$--$3$ are routinely needed to absorb the mismatch between $\e_C^n$ and the data~\cite{Cornella2023}.

Two clarifications frame the role of these coefficients. First, the deviations from unity are outputs of the comparison rather than inputs; the hypothesis is that the exponents lie on the lattice with unit coefficients, and the deviations $|c|\leq1.07$ across the seven observables of Table~\ref{tab:scalings} (Table~\ref{tab:lattice-fit}, row $d=9$) are its measured residuals. Deviations of this size are within the resolution at which the hypothesis can currently be tested, since the running-mass inputs carry parametric uncertainties at the few-percent level (larger for the first generation) and higher-order corrections of comparable size are expected; present data cannot distinguish $c=1$ exactly from $|c-1|$ of a few percent, and improved determinations can falsify the hypothesis. Second, a lattice of resolution $1/9$ guarantees that any single observable lies within a factor $\e^{1/18}\simeq0.91$ of some lattice point, so no individual residual is evidence by itself. The evidence is joint: the exponents follow from the additive charges of Table~\ref{tab:charges} rather than being chosen observable by observable; with $\e$ fixed by $|V_{us}|$, the relation $|V_{cb}|=\e^{17/9}$ holds to $0.4\%$ with no continuous freedom (Sec.~\ref{sec:Bvalue}); the charged-lepton ratios reuse the same $\e$ on the commensurate $1/6$ sublattice; and Table~\ref{tab:lattice-fit} quantifies the preference for the resolutions $1/9$ and $1/6$ over alternative denominators.

Worked examples showing how specific mass ratios and CKM elements emerge from the hop framework are collected in Appendix~\ref{app:examples}.

\subsection{Lepton exponent matrices}
\label{sec:lepton-textures}

Beyond quarks, the charged-lepton mass ratios are reproduced by the same expansion parameter, though on a finer $1/6$ lattice (the half-integer lepton-doublet charges place the charged leptons on sixths rather than ninths; see Sec.~\ref{sec:why-nine}),
\begin{align}
\frac{m_\mu}{m_\tau} &\sim \e^{5/3}, &
\frac{m_e}{m_\mu} &\sim \e^{19/6},
\label{eq:lepton-scalings}
\end{align}
as shown in Table~\ref{tab:scalings}. The charged-lepton exponent matrix, using left-handed doublet charges $Q(L_i) = (1, 1/2, 0)$ and right-handed charges $Q(e^c_j) = (23/6, 7/6, 0)$ from Table~\ref{tab:charges}, takes the form
\begin{equation}
p^{\ell}_{ij} =
{\setlength{\arraycolsep}{6pt}\renewcommand{\arraystretch}{1.15}
\frac{1}{6}\begin{pmatrix}
29 & 13 & 6\\
26 & 10 & 3\\
23 & 7  & 0
\end{pmatrix}}.
\label{eq:lepton-texture}
\end{equation}
For neutrinos, the effective Majorana mass matrix arises from the dimension-five Weinberg operator~\cite{WeinbergDim5} $LLHH/\Lambda_\nu$ and is symmetric in the lepton-doublet indices, giving exponents $p^\nu_{ij} = Q(L_i) + Q(L_j)$:
\begin{equation}
p^{\nu}_{ij} =
{\setlength{\arraycolsep}{6pt}\renewcommand{\arraystretch}{1.15}
\begin{pmatrix}
2 & 3/2 & 1\\
3/2 & 1 & 1/2\\
1 & 1/2  & 0
\end{pmatrix}}.
\label{eq:nu-texture}
\end{equation}
The mass eigenvalue hierarchy is $m_1:m_2:m_3 \sim \e^2:\e:1$, corresponding to normal ordering. Large atmospheric mixing ($\theta_{23}\sim45^\circ$) requires the $\mathcal{O}(1)$ coefficients in the $23$ block to exhibit approximate $\mu$--$\tau$ symmetry; this is a condition on the coefficients, not on the exponents. We note that the texture in Eq.~\eqref{eq:nu-texture} follows from strict charge additivity and therefore has a hierarchical $23$ block (entries $1$, $1/2$, $1/2$, $0$). In the companion lepton-lattice paper~\cite{LeptonLattice}, the effective neutrino texture instead adopts a flat $23$ block with $p^\nu_{22}=p^\nu_{23}=p^\nu_{33}=0$, enforced by a messenger selection rule that freezes the second- and third-generation lepton-doublet charges at the same effective value. The flat structure embeds $\mu$--$\tau$ symmetry directly in the exponents rather than relying on $\mathcal{O}(1)$ coefficients alone, and is responsible for the sharp octant--$\delta_{\rm CP}$ correlation found there. Both textures reproduce the same mass eigenvalue hierarchy; they differ in the mechanism for large $\theta_{23}$, and the choice between them is ultimately a question about the messenger sector in the lepton doublet channel. We return to the neutrino sector in Sec.~\ref{sec:neutrinos}.

The unification of quark and lepton textures on a single commensurate lattice (quark ninths and lepton sixths, both sublattices of the $1/18$ lattice) is one of the main phenomenological motivations for seeking a discrete gauge origin.
From the model-building perspective, the ninths quantization is unusually robust: adding heavy messenger chains shifts exponents by integers, but cannot spoil the underlying $1/9$ granularity. This makes the ninths lattice a clean discriminator between genuinely discrete-charge constructions and purely continuous FN models with arbitrary rational charges. In practice, the lattice pins down which texture deformations are allowed at a given order in $\e$, and this is what enables sharp statements about CKM angles and CP phases in the companion analyses~\cite{Companion,TwoOverTwo}.

\subsection{Renormalization-scale dependence}
\label{sec:RG-dependence}

The exponents $p_{ij}$ are fixed by the discrete charges and do not run; the entire renormalization-scale dependence of the Yukawa couplings resides in the coefficients $c_{ij}(\mu)\equiv Y_{ij}(\mu)/\e^{\,p_{ij}}$. The lattice hypothesis therefore separates into a scale-independent part (the exponent assignment) and a scale-anchored part (the clustering of the coefficients at unity). The anchor scale is the one stated in Table~\ref{tab:scalings}; the quark ratios are evaluated at $\mu\sim10^{2}$~GeV with the running masses of Ref.~\cite{HuangZhou}, and the CKM inputs are the low-energy measured values, the same choice made in integer-charge FN analyses~\cite{Cornella2023}.

The quantitative question is how far the coefficients drift between $M_Z$ and the messenger scale $\Lambda\simeq(3$--$4)\times10^{12}$~GeV at which the operators are generated. Over this range $|V_{us}|$ is stable to below one part in $10^{3}$ (only the third-row and third-column CKM elements run appreciably, and they do so by a common multiplicative factor), and the charged-lepton mass ratios are stable at the percent level. The scale sensitivity concentrates in the quantities involving the top Yukawa coupling: $m_c/m_t$, $m_s/m_b$, $|V_{cb}|$, and $|V_{ub}|$. Integrating the one-loop equations from $M_Z$ to $\Lambda$, the most sensitive ratio, $m_c/m_t$, decreases by roughly $10\%$ with Standard-Model running and by roughly $25\%$ with MSSM-like running at $\tan\beta\simeq16$; $m_s/m_b$, $|V_{cb}|$, and $|V_{ub}|$ shift at the $10\%$ level, with signs and coefficients that depend on the Higgs-sector completion. One unit of the $1/9$ lattice is a factor $\e^{1/9}\simeq0.83$, a step of $17$--$20\%$, so the integrated drift amounts to about half a lattice unit, at most about one and a half units for $m_c/m_t$ in the supersymmetric case, and every $c_{ij}(\mu)$ remains of order unity, within a factor of about $1.4$, at all scales between $M_Z$ and $\Lambda$. Nothing in the construction, which requires only $\mathcal{O}(1)$ coefficients, is destabilized by the running. What the running does select is the scale at which the coefficients cluster sharply at unity, $\mu\sim M_Z$; we regard the scale at which the relation sharpens as information about the relation, in the same sense that $b$--$\tau$ unification is stated at its matching scale (Appendix~\ref{app:examples}), rather than as an ambiguity of the framework.

\section{\texorpdfstring{$\bm{\Z_{18}}$}{Z18} Origin, Ninths Substructure, and Anomaly Consistency}
\label{sec:Z18}

\subsection{Ninths from \texorpdfstring{$\bm{\Z_9}$}{Z9}}

The ninths lattice suggests that FN charges are defined modulo nine. 
A minimal way to generate all residues is a set of flavon charges $(1,2,4)\,(\mathrm{mod}\,9)$, Fig.~\ref{fig:Z9clock}.

\begin{figure}[t]
\centering
\begin{tikzpicture}[scale=1.0]
\draw (0,0) circle (1.85cm);
\foreach \x in {0,...,8}
{
    \node at ({90-40*\x}:1.85cm) {\x};
}
\draw[->,thick] ({90}:1.35cm) -- ({50}:1.35cm);
\draw[->,thick] ({90}:1.05cm) -- ({10}:1.05cm);
\draw[->,thick] ({90}:0.75cm) -- ({-70}:0.75cm);
\end{tikzpicture}
\caption{$\Z_9$ flavor subgroup of $\Z_{18}$. The generators $(1,2,4)$ span all residues mod 9, enabling suppressions quantized in $1/9$.}
\label{fig:Z9clock}
\end{figure}

Equivalently, all effective FN charges may be written as
\begin{equation}
q_i=\frac{k_i}{9},\qquad k_i\in\Z,
\label{eq:qk9}
\end{equation}
and operator selection is the statement that total charge is an integer.
 
Yukawa operators obey discrete invariance conditions, e.g.
\begin{equation}
X(Q_i)+X(u_j)+X(H_u)\equiv0 \pmod{9},
\label{eq:yuk-inv}
\end{equation}
and similarly for down-type and lepton operators, enforcing that every FN suppression exponent lies in $\frac{1}{9}\Z$. 
In this sense, the continuous FN symmetry is an accidental low-energy remnant of a gauged $\Z_{18}$ selection rule. The structure is most transparent sector by sector. In the \emph{quark} sector the mod-9 selection rule [Eq.~\eqref{eq:yuk-inv}] forces every FN suppression exponent into $\tfrac{1}{9}\Z$: the quark mass ratios and CKM angles are all powers of $\e^{1/9}$, so the quark flavor lattice has spacing $1/9$ and is controlled by the $\Z_9$ subgroup. The \emph{charged-lepton} sector, however, lives on a finer $1/6$ lattice: the half-integer lepton-doublet charges $Q(L_i)=(1,\tfrac12,0)$ combine with the sixths in $Q(e^c_i)$ to give mass-ratio exponents $m_e/m_\tau\sim\e^{29/6}$ and $m_e/m_\mu\sim\e^{19/6}$, which are not in $\tfrac19\Z$. The two sectors therefore demand denominators $9$ and $6$, and the combined quark--lepton flavor lattice has spacing $1/18$, where $18=\mathrm{lcm}(9,6)$, with all exponents in $\tfrac1{18}\Z$.

This already fixes the discrete group from flavor data alone: the smallest cyclic group containing both a $\Z_9$ (quark ninths) and a $\Z_6$ (lepton sixths) is $\Z_{\mathrm{lcm}(9,6)}=\Z_{18}$. Equivalently $\Z_{18}\cong\Z_9\times\Z_2$ (since $\gcd(9,2)=1$), with the $\Z_9$ accommodating the quark sector and the extra $\Z_2$ supplying the factor of two needed for the charged-lepton sixths. This is the \emph{same} $\Z_{18}$ that the axion-quality argument requires for an independent reason; it is the double cover of the quark $\Z_9$, pushing the leading PQ-violating operator to dimension eighteen rather than nine (Sec.~\ref{sec:quality}). The $\Z_2$ extension thus does double duty: it accommodates the half-integer lepton charges in the flavor sector and protects axion quality. The order of the group is therefore overdetermined, fixed by the combined flavor data through $\mathrm{lcm}(9,6)=18$ and reinforced by the quality requirement.

The point of the $(1,2,4)$ choice is that it is \emph{minimal} in a strong sense: these charges, being $\{2^0,2^1,2^2\}$, correspond to $N_{\rm hop}=3$ distinct types of nearest-neighbor hop on a chain of $N_{\rm site}=4$ vectorlike messenger sites. (The chain sites carry $\Z_9$ charges $(0,8,6,2)$; the hop charges $(1,2,4)$ are the differences between adjacent sites, since $0-8\equiv1$, $8-6=2$, $6-2=4$ mod~9. Thus $N_{\rm hop}=N_{\rm site}-1$.) Every residue mod~9 can be reached as a non-negative integer combination of the three hop charges with total hop count $n_{\rm tot}\equiv n_1+n_2+n_3$ at most three, so that no FN exponent requires more than three flavon insertions. (Here $n_a$ denotes the number of hops of type $a$, each hop contributing charge $q_a/9$; $n_{\rm tot}$ is the total number of hops, which should not be confused with the fixed number of hop types $N_{\rm hop}=3$ or the number of chain sites $N_{\rm site}=4$.) No two-element subset of hop charges achieves this (e.g.\ the pair $(1,2)$ requires $n_{\rm tot}$ up to four). As a result, the allowed singlet monomials built from flavons with these charges satisfy the congruence
\begin{equation}
n_1 + 2n_2 + 4n_3 \equiv 0 \pmod{9},
\label{eq:congruence}
\end{equation}
which directly controls the spectrum of FN suppressions. Every residue mod~9 is realized by a minimal hop combination with $n_{\rm tot}\leq 3$, as listed in Table~\ref{tab:residues}; the residue $7$, attainable only via $(n_1,n_2,n_3)=(1,1,1)$, is what forces all three hop types and hence $n_{\rm tot}=3$.
\begin{table}[h]
\centering
\caption{Minimal hop content $(n_1,n_2,n_3)$ realizing each residue $r\equiv n_1+2n_2+4n_3\pmod 9$, with the associated FN suppression $\e^{r/9}$ and total hop count $n_{\rm tot}=n_1+n_2+n_3$. All nine residues are reached with $n_{\rm tot}\leq 3$.}
\label{tab:residues}
\renewcommand{\arraystretch}{1.15}
\begin{tabular}{cccc}
\hline\hline
$r$ & $(n_1,n_2,n_3)$ & $n_{\rm tot}$ & suppression \\
\hline
$0$ & $(0,0,0)$ & $0$ & $1$ \\
$1$ & $(1,0,0)$ & $1$ & $\e^{1/9}$ \\
$2$ & $(0,1,0)$ & $1$ & $\e^{2/9}$ \\
$3$ & $(1,1,0)$ & $2$ & $\e^{3/9}$ \\
$4$ & $(0,0,1)$ & $1$ & $\e^{4/9}$ \\
$5$ & $(1,0,1)$ & $2$ & $\e^{5/9}$ \\
$6$ & $(0,1,1)$ & $2$ & $\e^{6/9}$ \\
$7$ & $(1,1,1)$ & $3$ & $\e^{7/9}$ \\
$8$ & $(0,0,2)$ & $2$ & $\e^{8/9}$ \\
\hline\hline
\end{tabular}
\end{table}
The full $\Z_{18}$ imposes a stronger constraint on PQ-violating scalar operators: since the flavon carries the minimal $\Z_{18}$ charge, the first gauge-invariant pure-flavon monomial is $\Phi^{18}$, not $\Phi^{9}$.

\subsection{Charge assignments}
\label{sec:charges}

The Froggatt--Nielsen charges that reproduce the textures in Eqs.~\eqref{eq:Y-diag-def}--\eqref{eq:nu-texture} via the additive rule $p_{ij}=Q(\psi_i)+Q(\psi^c_j)$ are given in Table~\ref{tab:charges}. These are the charges established in the companion two-over-two paper~\cite{TwoOverTwo}; the lepton charges include the corrections from the companion lepton-lattice paper~\cite{LeptonLattice}. All charges lie on the $1/18$ sublattice (i.e.\ $18Q$ is an integer for every field), consistent with the $\Z_{18}$ discrete gauge symmetry. The embedding of these FN charges into a self-consistent set of $\Z_{18}$ integer charges, including the chain messenger sector, is developed in detail in the companion unified flavor paper~\cite{UFP}.

From the additive rule one finds, for example, that the down-type $(1,1)$ exponent is
\begin{equation}
p^d_{11} = Q(Q_1) + Q(d^c_1) = 3 + \tfrac{10}{9} = \tfrac{37}{9}\,,
\end{equation}
matching the $(1,1)$ entry of $p^d_{ij}$ in Eq.~\eqref{eq:Y-diag-def}. The unsuppressed $(3,3)$ entries follow from $Q(Q_3)=Q(u^c_3)=Q(d^c_3)=0$. The exponent \emph{differences} (which control mass ratios and mixing angles) are the physically meaningful quantities, and the charge assignments in Table~\ref{tab:charges} reproduce all seven entries of Table~\ref{tab:scalings} with $\mathcal{O}(1)$ coefficients close to unity~\cite{TwoOverTwo}.

\subsection{Anomaly cancellation}
\label{sec:anomaly}

If $\Z_{18}$ is a discrete gauge symmetry, the Iba\~nez--Ross/Banks--Dine conditions~\cite{IbanezRoss,BanksDine} are
\begin{align}
A_3 &\equiv \sum_{\rm color} q_i\,d_2(i)\equiv 0\pmod{18},\label{eq:anomaly}\\
A_2 &\equiv \sum_{\rm weak} q_i\,d_3(i)\equiv 0\pmod{18},\nonumber\\
A_{\rm grav} &\equiv \sum_{\rm all} q_i\,d_3(i)d_2(i)\equiv 0\pmod{18},\nonumber
\end{align}
where $q_i$ is the integer $\Z_{18}$ charge of the $i$th left-handed Weyl fermion and $d_2,d_3$ are the $SU(2)_L$ and $SU(3)_C$ representation dimensions. (We adopt the convention~\cite{IbanezRoss,BanksDine} in which all three conditions are stated $\pmod{18}$; the alternative gravitational convention $\pmod{N/2}$ for even $N$ is more permissive and is discussed below.) The Higgs sector is two-Higgs-doublet (DFSZ-II): $H_u$ gives mass to up-type quarks and $H_d$ to down-type quarks and charged leptons. The Higgses are scalar fields and contribute only indirectly through the Yukawa-invariance constraints; the sums thus run over the Standard-Model fermion content plus right-handed neutrinos, matching the convention of the companion unified-flavor analysis~\cite{UFP}. (In the supersymmetric realization the higgsinos, with $X(\tilde H_u)+X(\tilde H_d)=-2$, would add $\Delta A_2=-2$ and $\Delta A_{\rm grav}=-4$, shifting the residues to $A_2\equiv7$ and $A_{\rm grav}\equiv8\pmod{18}$ without altering the conclusion that the residual anomalies are nonzero and must be cancelled in the UV.) Using the integer charges $q_{18}\equiv 18\,Q$ obtained from Table~\ref{tab:charges} (and consistent with the lepton charges of Ref.~\cite{LeptonLattice}), the SM matter content of three generations gives the residues
\begin{equation}
(A_3,A_2,A_{\rm grav})\equiv(16,9,12)\pmod{18}\,,
\label{eq:anomaly-residues}
\end{equation}
all non-zero. The SM matter charges fixed by the observed flavor texture therefore do \emph{not} by themselves satisfy the discrete anomaly conditions. This is a generic feature of FN constructions and requires UV input; the present subsection clarifies what is assumed and what is derived.

\paragraph{Why standard universal Green--Schwarz cancellation fails.}
A continuous-$U(1)$ Green--Schwarz mechanism can absorb mixed anomalies only when they are \emph{universal}, in the sense that a single axionic shift cancels all of them simultaneously. With Kac--Moody levels $(k_3,k_2)=(1,1)$ for $SU(3)_C$ and $SU(2)_L$ in the canonical $SU(5)$-GUT normalization, the universality condition reduces to
\begin{equation}
A_3 \;=\; A_2 \pmod{18}\,,
\end{equation}
which is \emph{not} satisfied by the residues \eqref{eq:anomaly-residues}: $16\not\equiv 9\pmod{18}$, with difference $7$ that is not a multiple of $18$. The residues therefore cannot be cancelled by a single axion shift acting universally on $G\widetilde G$ and $W\widetilde W$, and a strict stringy GS mechanism with one universal axion does not by itself complete the construction.

\paragraph{What the construction does require.}
The discrete anomaly residues must be removed by genuine UV input. We list the available options as they are organized in the companion paper~\cite{UFP}:
\begin{enumerate}
\item \emph{Spectator fermions.} A set of SM-singlet (or otherwise UV-localized) heavy fermions carrying $\Z_{18}$ charges chosen to bring Eq.~\eqref{eq:anomaly-residues} to $(0,0,0)\pmod{18}$. The smallest viable set adds three Weyl singlets with $\Z_{18}$ charges that match $(-A_3,-A_2,-A_{\rm grav})$ component by component; this is the option developed in Ref.~\cite{UFP}.
\item \emph{Two-axion non-universal Green--Schwarz mechanism.} Generalizations of GS exist in which several axion-like fields cancel non-universal anomalies, as in heterotic compactifications with multiple model-independent axions. Such a multi-axion completion would require additional pseudoscalar moduli not present in the IR construction described here.
\item \emph{Anomaly-free remnant interpretation.} If the underlying $U(1)_{\rm FN}$ is regarded as a global flavor symmetry (rather than an exact gauge symmetry) that is broken at high scale, then the discrete remnant $\Z_{18}$ is anomaly-free as an accidental symmetry of the IR theory, even though the parent $U(1)_{\rm FN}$ is anomalous. In this interpretation the Banks--Dine conditions \eqref{eq:anomaly} are not strictly required, and the residues \eqref{eq:anomaly-residues} parametrize the explicit breaking of the global $\Z_{18}$ by Planck-suppressed operators of the form discussed in Sec.~\ref{sec:quality-op}.
\end{enumerate}
Of these, option (1) is the cleanest as a discrete-gauge construction and is the one developed in detail in the UV-complete companion paper~\cite{UFP}. Option (3) is sufficient for the IR phenomenology of the present paper, which depends only on the accidental $U(1)_{\rm PQ}$ at low energy and on the leading Planck-suppressed PQ-violating operator $\Phi^{18}$, both of which are insensitive to whether $\Z_{18}$ is exact gauge or accidental global. Option (2) is in principle viable but requires additional UV ingredients not present in the minimal construction.

\paragraph{Convention on the gravitational anomaly.}
Some authors~\cite{BanksDine} take the gravitational discrete anomaly condition with reduced modulus, $A_{\rm grav}\equiv 0\pmod{N/2}$ for even $N$, on the ground that no Majorana mass terms are present in the SM. In this convention $A_{\rm grav}\equiv 12\equiv 3\pmod{9}$, which is also non-zero. Switching conventions therefore does not relax the cancellation requirement; the residue is genuine.

\paragraph{What this paper assumes.}
For the IR predictions of this paper (the dimension-eighteen leading PQ-violating operator, the quality estimate of Sec.~\ref{sec:quality}, the light-spectrum anomaly ratio $E/N\simeq2.6$--$3.0$, and the dark-matter mass window), only two pieces of information are needed: (a) that $\Z_{18}$ acts on the matter sector with the integer charges of Table~\ref{tab:charges}, and (b) that the underlying UV theory exists and is consistent. Two quantities do retain completion dependence: the integer domain-wall number [Eq.~\eqref{eq:ndw-dfsz}], for which we adopt the canonical benchmark $N_{\rm DW}=6$; and possible additive heavy-sector contributions to $E$ and $N$, since $\Z_{18}$-chiral spectators with flavon-induced masses carry PQ charge and anomalous contributions do not decouple with mass. Throughout we quote the light-spectrum (Standard Model plus higgsino) anomaly ratio as the irreducible matter-sector prediction of the $\Z_{18}$ charge assignment. With this scoping, whether (b) is realized via spectator fermions (option 1), a multi-axion GS mechanism (option 2), or as an accidental global symmetry of the IR theory (option 3) does not affect the IR phenomenology developed in Secs.~\ref{sec:quality}--\ref{sec:pheno}. We therefore proceed with the IR analysis under the assumption that one of the above completions is realized (most concretely the spectator-fermion completion of Ref.~\cite{UFP}), and note explicitly that the GS-style language used elsewhere in this paper refers to the abstract role of an axion in absorbing residual anomalies, not to a strict universal-GS mechanism.

The chain VLQs of Sec.~\ref{sec:messengers} are vectorlike under $\Z_{18}$ in the assignment of Ref.~\cite{UFP} and therefore do not contribute to the anomaly sums; the residues in Eq.~\eqref{eq:anomaly-residues} arise entirely from the SM fermion charges fixed by the flavor texture. The explicit numerical computation of $(A_3,A_2,A_{\rm grav})$ is given in Appendix~\ref{app:anomaly}.

In a minimal UV completion, the chain has $N_{\rm site}=4$ sites carrying $\Z_9$ charges $(0,8,6,2)$, connected by $N_{\rm hop}=3$ links with hop charges $(1,2,4)$. All four sites are new heavy vectorlike messenger pairs $D_a+\bar{D}_a$ ($a=1,2,3,4$) with masses $M_a\sim\Lambda$; the SM right-handed quarks $d_{R,j}$ couple to $D_1$ and the SM left-handed doublets $q_{L,i}$ couple to $D_4$ via endpoint Yukawa operators~\cite{UFP}. Integrating out all four VLQ pairs at tree level generates the FN operators in Eq.~\eqref{eq:FN-op}, with the number of hops $n_a$ of each type determining the suppression exponent.

\section{Axion Quality and the Domain-Wall Number}
\label{sec:quality}

We now identify the FN flavon with the PQ field, so that $\Phi$ carries the minimal $\Z_{18}$ charge $q_\Phi=1/18$, and the angular mode of $\Phi$ is the QCD axion. The axion sector is realized through a two-Higgs-doublet structure of standard ``Type-II'' form~\cite{DiLuzioReview}: up-type quarks couple to $H_u$, while down-type quarks and charged leptons couple to $H_d$, exactly as in the supersymmetric (Minimal Supersymmetric Standard Model, MSSM-like) two-Higgs-doublet model. We denote the resulting axion ``DFSZ-II,'' following the companion analysis~\cite{Subconstituents}, where ``Type-II'' refers throughout to this two-Higgs-doublet classification (each of the up- and down-type fermions coupling to a separate doublet), which fixes $E/N=8/3$ in the universal-charge limit. We note that the labels ``DFSZ-I'' and ``DFSZ-II'' are not used uniformly in the axion literature, and we have specified the Higgs assignment explicitly above to avoid ambiguity. This is the same Higgs structure that, combined with the chain internal factor, fixes $\tan\beta\simeq 10$--$16$~\cite{Subconstituents}. Throughout this section we establish two structural results: the leading Planck-suppressed PQ-violating operator is the dimension-eighteen monomial $\Phi^{18}$ (Sec.~\ref{sec:quality-op}), so that axion quality is automatic, and the canonical axion field and physical decay constant take the form of Sec.~\ref{sec:axion-norm}. The domain-wall number, fixed by the color anomaly of the two-Higgs-doublet sector, is the canonical three-generation DFSZ-II value $N_{\rm DW}=6$ (Sec.~\ref{sec:NDW-DFSZ}).

\subsection{Leading Planck-suppressed operator}
\label{sec:quality-op}

The claim that the leading Planck-suppressed PQ-violating operator is exactly $\Phi^{18}$ rests on three structural inputs: (i) the scalar content of the model, (ii) the $\Z_{18}$ charges of all fields, and (iii) the absence of GUT-breaking fields with nontrivial $\Z_{18}$ charge. We make these explicit.

The scalar content relevant to PQ-violating operators is the single flavon $\Phi$ together with the two Higgs doublets $H_u,H_d$; there are no other $\Z_{18}$-charged scalars below the GUT scale. We enumerate the gauge-invariant operators these can form and show that the first PQ-violating one appears at dimension eighteen.

\paragraph{Scalar content.}
The scalar sector of the model contains exactly the following gauge-invariant building blocks, with their $\Z_{18}$ charges (quoted in units of $1/18$, i.e.\ as integers in $\Z_{18}$). In the DFSZ-II assignment the two Higgs doublets carry PQ charges summing to $X(H_u)+X(H_d)\equiv H=-2$ (Sec.~\ref{sec:NDW-DFSZ}), fixed by Yukawa invariance relative to the flavon:
\begin{equation}
\renewcommand{\arraystretch}{1.15}
\begin{array}{l|c|l}
\hline
\text{Field} & q\in\Z_{18} & \text{role} \\
\hline
\Phi & +1 & \text{single flavon (PQ field)} \\
\Phi^\dagger & -1 & \text{conjugate flavon} \\
H_u H_d & -2 & \text{$\mu$-term Higgs bilinear} \\
H_u^\dagger H_u,\,H_d^\dagger H_d & 0 & \text{neutral Higgs bilinears} \\
\hline
\end{array}
\nonumber
\end{equation}
There is \emph{one} flavon $\Phi$, not several. The "hop charges" $(1,2,4)$ of Sec.~\ref{sec:Z18} arise from \emph{differences of fermion-site $\Z_{18}$ charges} along the messenger chain, not from distinct scalar flavons: every flavon insertion in an FN operator is the same $\Phi$, but the chain topology selects which combinations of insertions are tree-level-generated by integrating out the messenger fermions. The messenger fermions $D_a+\bar D_a$ and $U_a+\bar U_a$ are fermions, not scalars, and contribute to PQ-violating scalar operators only through their fermion bilinears $\bar\psi\chi$ which carry electroweak quantum numbers and dimension three, suppressed by powers of the heavy mass and the Higgs VEV.

GUT-breaking scalars (the $SU(5)$-breaking adjoint $\bm{24}$, the $SO(10)$-breaking $\bm{210}$ or adjoint, and the doublet--triplet-splitting Higgs $\bm{45}+\overline{\bm{45}}$ if present) are taken to be $\Z_{18}$-singlets, in line with the standard model-building convention that GUT-breaking dynamics decouple from the flavor and PQ sectors. With this assumption, GUT-breaking VEVs cannot construct $\Z_{18}$-violating operators below dimension eighteen.

\paragraph{Operator enumeration up to dimension eighteen.}
A general non-derivative gauge-invariant operator built from the scalar fields has the form
\begin{equation}
\mathcal{O} \;=\; \Phi^{a}(\Phi^\dagger)^{b}\,(H_u H_d)^{c}\,(H_u^\dagger H_u)^{d_u}(H_d^\dagger H_d)^{d_d},
\label{eq:gen-op}
\end{equation}
where $a,b,c,d_u,d_d\geq 0$, with mass dimension $a+b+2c+2d_u+2d_d$ and $\Z_{18}$ charge $a-b-2c$ (since $H_u H_d$ carries charge $-2$ and the neutral bilinears carry zero). PQ-violation requires $\mathcal{O}$ to be \emph{not} expressible as a polynomial in the PQ-invariant building blocks $|\Phi|^2$, $H_u^\dagger H_u$, $H_d^\dagger H_d$, and the $\mu$-block $H_u H_d\,\Phi^2$ (which is $\Z_{18}$-neutral). $\Z_{18}$-invariance requires
\begin{equation}
a-b-2c\;\equiv\;0\pmod{18}\,.
\end{equation}
For a genuinely PQ-violating holomorphic operator (one not equal to a power of the $\mu$-block times $|\Phi|^2$), the smallest solution is the pure-flavon monomial $(a,b,c)=(18,0,0)$ of dimension eighteen.

What about \emph{mixed} monomials involving Higgs bilinears? These are enumerated in Table~\ref{tab:dangerous-operators}. The only $\Z_{18}$-charged scalar monomials available below dimension eighteen are $\Phi^{n}$ and $(\Phi^\dagger)^{n}$ with $n\leq 17$ (charge $\pm n$), dressed by the Higgs bilinear $H_u H_d$ (charge $-2$) or the neutral bilinears (charge $0$). A holomorphic operator $\Phi^{n}(H_u H_d)^{m}$ has $\Z_{18}$ charge $n-2m$; setting this to zero requires $n=2m$, which is the $\mu$-block raised to the $m$th power and is therefore PQ-conserving. Any \emph{genuinely} PQ-violating operator must instead have $n-2m=18$, whose lowest-dimension representative is again $\Phi^{18}$ (dimension eighteen); the next, $\Phi^{20}(H_u H_d)$, has dimension twenty-two. No combination of the available Higgs bilinears reaches a PQ-violating, $\Z_{18}$-invariant operator below dimension eighteen.

\begin{table*}[t]
\centering
\caption{Enumeration of all $\Z_{18}$-invariant operator candidates up to dimension eighteen, classified by whether they violate the accidental $U(1)_{\rm PQ}$. Dimensions in parentheses count the mass dimension of each building block (3 for fermion bilinears, 1 for scalars). Operators in the lower section are PQ-conserving and pose no quality threat. The first PQ-violating operator is $\Phi^{18}$ at dimension eighteen.}
\label{tab:dangerous-operators}
\renewcommand{\arraystretch}{1.2}
\begin{tabular}{lccl}
\hline\hline
Operator & Mass dim. & $\Z_{18}$ charge & PQ-violating? \\
\hline
$\Phi^{n}$, $1\leq n\leq 17$            & $n$  & $n$    & forbidden by $\Z_{18}$ \\
$\Phi^{n}(H_u H_d)$, $1\leq n\leq 15$, $n\neq 2$   & $n+2$  & $n-2$    & forbidden by $\Z_{18}$ \\
$\Phi^{n}(H_u^\dagger H_u)$, $1\leq n\leq 16$ & $n+2$  & $n$ & forbidden by $\Z_{18}$ \\
$\Phi^{n}(H_u H_d)^{2}$, $1\leq n\leq 13$, $n\neq 4$ & $n+4$  & $n-4$  & forbidden by $\Z_{18}$ \\
$\Phi^{18}$                              & $18$  & $0$    & \textbf{leading PQ-violating operator} \\
\hline
$|\Phi|^2$, $|\Phi|^{2k}$ ($k\geq 1$)         & $2k$  & $0$    & PQ-conserving (singlet) \\
$H_u^\dagger H_u$, $H_d^\dagger H_d$ & $2$ & $0$ & PQ-conserving \\
$H_u H_d\,\Phi^2$ ($\mu$-term, $n=2$) & $4$ & $0$ & PQ-conserving \\
\hline\hline
\end{tabular}
\end{table*}

\paragraph{Conclusion.}
The leading Planck-suppressed PQ-violating operator is therefore
\begin{equation}
\Delta \mathcal{L} \supset \frac{c_{18}}{\mpl^{14}}\,\Phi^{18} + \mathrm{h.c.},
\label{eq:planck18}
\end{equation}
with $c_{18}=|c_{18}|\,e^{i\delta}$ a complex Wilson coefficient of order unity. Writing $\Phi=(v_\Phi/\sqrt{2})\,e^{ia/v_\Phi}$ with $v_\Phi=N_{\rm DW}\,f_a=6\,f_a$ (Sec.~\ref{sec:axion-norm}), the operator generates a PQ-violating contribution to the axion potential
\begin{equation}
V_{\rm PQ}(a)\;=\;\frac{|c_{18}|}{2^{8}}\,\frac{v_\Phi^{18}}{\mpl^{14}}\,\cos\!\Bigl(\frac{18\,a}{v_\Phi}+\delta\Bigr)\,,
\label{eq:V-PQ-ampl}
\end{equation}
with amplitude scale $V_{\rm PQ}^{\rm amp}\equiv|c_{18}|v_\Phi^{18}/(2^{8}\mpl^{14})$. The QCD-induced potential is $V_{\rm QCD}(a)=-\chi_{\rm top}\cos(a/f_a)$, with topological susceptibility $\chi_{\rm top}=m_\pi^2 f_\pi^2\,m_u m_d/(m_u+m_d)^2\simeq(75.5\,\mathrm{MeV})^4\simeq 3.4\times 10^{-5}\,\mathrm{GeV}^4$. Minimising the sum, the shift of the axion vacuum from $\bar\theta=0$ is
\begin{equation}
\Delta\bar\theta\;\simeq\;\frac{18\,|c_{18}|\,\sin\delta}{2^{8}\,N_{\rm DW}}\,\frac{v_\Phi^{18}}{\mpl^{14}\,\chi_{\rm top}}\,,
\label{eq:dtheta}
\end{equation}
where the factor $18/N_{\rm DW}=18\,f_a/v_\Phi$ arises because the operator's phase advances as $18\,a/v_\Phi=(18/N_{\rm DW})\,\bar\theta$, and the linear approximation in $\Delta\bar\theta$ is justified throughout the regime of interest because $V_{\rm PQ}^{\rm amp}\ll\chi_{\rm top}$ by an enormous factor (see below).

For $|c_{18}|=1$, $|\sin\delta|=1$, and $f_a$ in the dark-matter window of Sec.~\ref{sec:pheno} (so $v_\Phi=6f_a$ for the benchmark domain-wall number $N_{\rm DW}=6$ of Sec.~\ref{sec:NDW-DFSZ}), using the \emph{reduced} Planck mass $\mpl=2.435\times 10^{18}$~GeV, Eq.~\eqref{eq:dtheta} gives
\begin{align}
f_a&=5\times 10^{11}\,\mathrm{GeV}: & |\Delta\bar\theta|&\simeq 5\times 10^{-31}\,, \nonumber\\
f_a&=7\times 10^{11}\,\mathrm{GeV}: & |\Delta\bar\theta|&\simeq 2\times 10^{-28}\,, \nonumber\\
f_a&=8\times 10^{11}\,\mathrm{GeV}: & |\Delta\bar\theta|&\simeq 3\times 10^{-27}\,,
\label{eq:dtheta-numerics}
\end{align}
even at the upper end of the predicted dark-matter window. (Relative to an $N_{\rm DW}=1$ realization at the same $f_a$, the net enhancement is $N_{\rm DW}^{17}=6^{17}\simeq 2\times10^{13}$: the amplitude grows as $v_\Phi^{18}=6^{18}f_a^{18}\simeq 10^{14}f_a^{18}$, while the slope in $\bar\theta$ carries one compensating factor $f_a/v_\Phi$. The bound remains satisfied by a wide margin.) With the full Planck mass $M_{\rm Pl}=1.22\times 10^{19}$~GeV the estimates are smaller by an additional factor $(M_{\rm Pl}/\mpl)^{14}\simeq 6\times 10^{9}$. In either convention, $|\Delta\bar\theta|$ is roughly sixteen orders of magnitude below the experimental bound $|\bar\theta|\lesssim 10^{-10}$ from neutron electric-dipole-moment searches. Even allowing $|c_{18}|$ as large as $\mathcal{O}(10^{16})$ (far beyond any plausible UV theory) would leave the bound satisfied; the construction provides an enormous safety margin against all order-one Wilson-coefficient uncertainties.

This is to be contrasted with generic DFSZ-II and KSVZ (Kim--Shifman--Vainshtein--Zakharov)~\cite{Kim,SVZ,DFSZ} constructions, where the leading PQ-breaking operator is typically allowed already at relatively low dimension ($n\leq 10$) unless additional ad hoc symmetries are imposed~\cite{KamionkowskiMarchRussell,Holman,BarrSeckel,BabuGogoladzeWang,BhattiproluMartin,BaerBargerSengupta}. In a model with only a $\Z_N$ symmetry protecting PQ, the leading operator $\Phi^N$ appears at dimension $N$; maintaining $|\Delta\bar\theta|<10^{-10}$ requires $N\geq 12$ for $f_a\sim 10^{12}$~GeV~\cite{KamionkowskiMarchRussell,Holman,BarrSeckel}. In the present framework, the prohibition of all operators $\Phi^n$ with $n<18$ is a direct consequence of the same $\Z_{18}$ arithmetic whose $\Z_9$ subgroup enforces the ninths flavor lattice, so axion quality is not an extra assumption but a corollary of the flavor structure, with the generous margin of safety quantified above.

We regard this as the central result of the paper, and it deserves emphasis. In essentially every other high-quality-axion construction, the protective symmetry is \emph{postulated} for the express purpose of forbidding the dangerous operators: one chooses a $\Z_N$ with $N$ large enough, or arranges a particular gauge or string structure, so that $\Phi^{n<N}$ is absent. Here the protective symmetry is not chosen at all. The discrete group $\Z_{18}$ is fixed, with no reference to the axion, by the requirement that the observed quark and lepton mass ratios and mixing angles be reproduced by exponents quantized in ninths (Secs.~\ref{sec:textures}--\ref{sec:Z18}): the denominator nine forces a $\Z_9$ selection rule, and the minimal flavon charge that generates the full ninths spectrum promotes it to $\Z_{18}$. Axion quality then follows as an unavoidable arithmetic consequence; the same charge $X(\Phi)=1$ that builds the $1/9$-quantized Yukawa textures forbids every PQ-violating monomial below $\Phi^{18}$. The quality of the axion is therefore \emph{predicted by flavor data}, not engineered: the dimension-eighteen suppression, and the resulting $|\Delta\bar\theta|\lesssim 3\times10^{-27}$ that sits roughly sixteen orders of magnitude below the neutron-EDM bound, are corollaries of a discrete group whose existence and order were fixed before the Peccei--Quinn mechanism was even introduced. The closest constructions in spirit are those of Sheng, Yanagida, and collaborators~\cite{ShengYanagida,GeorisYanagida,ShengYanagidaZhang}, who likewise protect the axion with a discrete gauge symmetry ($\Z_4\times\Z_3$) motivated by Standard-Model structure; the distinction here is that the group order ($N=18$) and the protective operator dimension are fixed \emph{quantitatively} by the measured fermion mass ratios and mixing angles through the ninths lattice, rather than by a qualitative appeal to the Standard-Model field content. This logic rests only on the scalar content $\Phi,H_u,H_d$ and the $\Z_{18}$ charges, independently of the detailed Higgs PQ-charge assignment.

\paragraph{Identification of the axion with the flavon phase.}
Since there is exactly one $\Z_{18}$-charged scalar field $\Phi$ that develops a large VEV (the Higgs doublets carry only the small charges fixed by Yukawa invariance, $X(H_u)+X(H_d)=-2$ in units of $1/18$), the angular mode of $\Phi$ is the unique pseudo-Goldstone boson of the spontaneously broken $\Z_{18}$. The axion of the present construction is identified with this angular mode: there is no axion mixing in the conventional sense, since $\theta_\Phi\equiv\arg\Phi$ is the dominant axion direction, with two-doublet mixing suppressed by $v_{u,d}^2/v_\Phi^2\sim 10^{-20}$ (Sec.~\ref{sec:axion-norm}). The QCD anomaly of the accidental $U(1)_{\rm PQ}$ is generated by the two-Higgs-doublet sector (Sec.~\ref{sec:NDW-DFSZ}); the discrete-anomaly residues~\eqref{eq:anomaly-residues} of $\Z_{18}$ on the SM matter sector are absorbed by the UV completion (Sec.~\ref{sec:anomaly}), most concretely by spectator fermions in the construction of Ref.~\cite{UFP}, and are not the responsibility of the IR axion field discussed here.

Since the flavon \emph{is} the PQ field, the flavor expansion parameter is directly $\e=\fa/\Lambda$, linking the messenger scale to the axion decay constant: $\Lambda=\fa/\e\sim(3$--$4)\times 10^{12}$~GeV for $\fa\sim(5$--$8)\times 10^{11}$~GeV. The flavor hierarchy and the axion mass window thus trace back to a single dimensionless ratio.

\subsection{Canonical axion field and physical decay constant}
\label{sec:axion-norm}

Writing the flavon as $\Phi(x)=\bigl[v_\Phi+\rho(x)\bigr]\,e^{i\,a(x)/v_\Phi}/\sqrt{2}$, with $v_\Phi=N_{\rm DW}f_a$ the radial scale that sets the periodicity of the angular field (the same $v_\Phi$ appearing in Eq.~\eqref{eq:V-PQ-ampl}), the angular mode $a(x)$ has canonical kinetic term and shifts under a $U(1)_{\rm PQ}$ rotation $\Phi\to e^{i\alpha q_\Phi}\Phi$ as
\begin{equation}
a(x)\,\to\,a(x)+\alpha\,q_\Phi\,v_\Phi,
\label{eq:a-shift}
\end{equation}
where $q_\Phi$ is the PQ charge of $\Phi$ in continuous-$U(1)_{\rm PQ}$ normalization. In the present construction the would-be PQ symmetry is the continuous lift of the discrete $\Z_{18}$, and the natural normalization is $q_\Phi=1$, since $\Phi$ carries the smallest $\Z_{18}$ charge. The two Higgs doublets carry small PQ charges $q_{H_u},q_{H_d}\in\frac{1}{18}\Z$ fixed by the Yukawa invariance conditions of Eq.~\eqref{eq:yuk-inv}. Because $v_\Phi\sim 10^{12}\,\mathrm{GeV}\gg v_u,v_d\sim 100\,\mathrm{GeV}$, the physical axion is, to one part in $10^{20}$, the angular mode of $\Phi$ alone:
\begin{equation}
a(x)\;\simeq\;v_\Phi\,\theta_\Phi(x)\,,
\qquad
\theta_\Phi\equiv\arg\Phi\,.
\label{eq:a-as-flavon}
\end{equation}
The Higgs admixture is proportional to $(v_u^2+v_d^2)/v_\Phi^2$ and is irrelevant for the axion phenomenology.

The QCD anomaly of the accidental $U(1)_{\rm PQ}$ is computed from the colored-fermion spectrum (Sec.~\ref{sec:NDW-DFSZ} and Appendix~\ref{app:EN-DFSZ}). The chain VLQs $D_a+\bar D_a$ are vectorlike under both the SM gauge group and $\Z_{18}$~\cite{UFP} and contribute zero; the anomaly is carried by the SM quarks through the two-Higgs-doublet PQ charges, giving $N=3$ and hence the domain-wall number
\begin{equation}
N_{\rm DW}=|2N|=6.
\end{equation}
As in the companion subconstituent analysis~\cite{Subconstituents}, the physical decay constant is identified directly with the flavon VEV,
\begin{equation}
\mathcal{L}\supset\frac{a(x)}{f_a}\,\frac{\alpha_s}{8\pi}\,G^{a\mu\nu}\widetilde G^a_{\mu\nu}\,,
\qquad
\boxed{f_a\;=\;\langle\Phi\rangle\,,}
\label{eq:fa-def}
\end{equation}
so that the radial scale $v_\Phi=N_{\rm DW}f_a=6f_a$, a normalization specific to the present paper, sets the periodicity of the angular field while $f_a$ is the scale appearing in the canonically normalized $a\,G\widetilde G$ coupling. The flavor expansion parameter is fixed by the flavon VEV through $\e=f_a/\Lambda$, so that $\Lambda=f_a/\e\simeq 5.4\,f_a$ given the data-determined $\e=14/75$. For $f_a\in(5$--$8)\times10^{11}$~GeV, this places the messenger scale at $\Lambda\simeq(3$--$4)\times 10^{12}$~GeV, the same scale that sets the seesaw Majorana mass for neutrinos (Sec.~\ref{sec:neutrinos}).

The QCD anomaly of $U(1)_{\rm PQ}$ relates the axion mass to $f_a$ via the standard chiral-Lagrangian result
\begin{equation}
m_a\,f_a\;=\;m_\pi\,f_\pi\,\frac{\sqrt{m_u m_d}}{m_u+m_d}\,,
\label{eq:ma-fa}
\end{equation}
which gives $m_a\simeq 5.7\,(10^{12}\,\mathrm{GeV}/f_a)\,\mu\mathrm{eV}$. For the predicted $f_a$ window this yields $m_a\simeq 7$--$12\,\mu\mathrm{eV}$.

Since $N_{\rm DW}=6$, the post-inflationary breaking of $\Z_{18}$ would produce a stable string--wall network; cosmological viability therefore requires that the PQ transition occur \emph{before} (or during) inflation, so that the network is inflated away and the isocurvature bound on the inflationary scale is respected. This is the standard requirement for $N_{\rm DW}>1$ axion models~\cite{Sikivie1982}, and is the cosmological history adopted in the companion subconstituent analysis~\cite{Subconstituents}, where the same single scale $\Lambda$ organizes the flavor, axion, and neutrino sectors.

\subsection{Domain-wall number: \texorpdfstring{$\bm{N_{\rm DW}=6}$}{NDW=6}}
\label{sec:NDW-DFSZ}

The QCD anomaly is carried by the Standard-Model quarks through the two-Higgs-doublet sector, with $H_u$ and $H_d$ carrying PQ charges fixed by Yukawa invariance, $X(H_u)+X(H_d)\equiv H=-2$ (in units of the fundamental $\Z_{18}$ charge $q_\Phi=1$). In the \emph{universal-charge limit} ($p^f_{ii}\to0$), the conditions $X(u^c_i)=-X(Q_i)-X(H_u)$ and $X(d^c_i)=-X(Q_i)-X(H_d)$ make the matter charges cancel, and in the standard color-anomaly normalization with $T(\mathbf 3)=1/2$ (the $SU(3)$ Dynkin index) and the weak-doublet multiplicity counted explicitly (Appendix~\ref{app:EN-DFSZ}), each quark generation contributes $2X(Q_i)+X(u^c_i)+X(d^c_i)=-H$, so
\begin{equation}
N_{\rm univ} \;=\; -\tfrac{3}{2}\,H \;=\; 3 ,
\label{eq:NDW-N}
\end{equation}
giving the canonical domain-wall number of a three-generation DFSZ-II axion~\cite{DiLuzioReview,Sikivie1982},
\begin{equation}
N_{\rm DW}\;=\;|2N_{\rm univ}|\;=\;6
\label{eq:ndw-dfsz}
\end{equation}
($N_{\rm DW}=2N_g$ with $N_g=3$). The chain vectorlike quarks $D_a+\bar D_a$ are $\Z_{18}$-vectorlike and contribute nothing~\cite{UFP}.

With the generation-\emph{dependent} charges of Table~\ref{tab:charges}, the matter contribution no longer cancels [Eq.~\eqref{eq:PQ-invariance}]: the color coefficient becomes $N=-119/18$ (Sec.~\ref{sec:pheno}), so that $|2N|=119/9$ is not an integer in flavon-charge units. The non-integer remainder is the uncancelled discrete color anomaly $A_3\equiv16\pmod{18}$ of Eq.~\eqref{eq:anomaly-residues}: $\Z_{18}$ invariance of the QCD-induced axion potential is restored only by the UV completion, so the physical (integer) domain-wall number is fixed by the completion rather than by the light matter content alone~\cite{UFP}; the spectator fermions that cancel the discrete anomaly acquire flavon-induced masses and hence carry PQ charge. Completions realizing the canonical value exist, and we adopt $N_{\rm DW}=6$ as the benchmark throughout. Two statements are completion-independent: $N_{\rm DW}>1$, so the post-inflationary string--wall network would be stable and the construction requires PQ breaking before or during inflation (Sec.~\ref{sec:axion-norm}); and the conclusion of the quality estimate, since Eq.~\eqref{eq:dtheta-numerics} rescales by $(N_{\rm DW}/6)^{17}$ and remains $\lesssim10^{-18}\ll10^{-10}$ even for $N_{\rm DW}=18$.

\section{\texorpdfstring{$\bm{E/N}$}{E/N} and Axion Phenomenology}
\label{sec:pheno}

The axion mass and its couplings to photons and matter depend on the electromagnetic and color anomaly coefficients $E$ and $N$. The color coefficient $N$, and hence the domain-wall number, was computed in Sec.~\ref{sec:quality}; here we derive the electromagnetic coefficient $E$ and the ratio $E/N$, which controls the axion--photon coupling. We then collect the axion mass relation, the cosmological relic density, and the lepton-flavor-violating couplings.

The electromagnetic anomaly coefficient is
\begin{equation}
E\;=\;\sum_i X_i\,d_i\,Q_i^2\,,
\label{eq:Edef}
\end{equation}
where $X_i$ is the chiral PQ charge of the $i$th left-handed Weyl fermion, $Q_i$ its electric charge, and $d_i$ the multiplicity from any non-Abelian internal indices summed over. The axion--photon coupling is
\begin{equation}
g_{a\gamma\gamma}=\frac{\alpha}{2\pi \fa}\left(\frac{E}{N}-2.0\right),
\label{eq:gagg}
\end{equation}
with $\alpha$ the fine-structure constant and the constant $2.0$ the model-independent QCD contribution, corresponding to the current PDG quark-mass ratio $m_u/m_d=0.47$~\cite{PDG2024}; we write $C_{a\gamma}\equiv|E/N-2.0|$. (The older value $1.92$ used in much of the literature corresponds to $m_u/m_d\simeq 0.60$.)\footnote{At next-to-leading order, the determination at physical quark masses gives $E/N-1.92(4)$~\cite{GrilliDiCortona}. We adopt the leading-order value $2.0$ at the PDG mass ratio for consistency with the companion analyses~\cite{Subconstituents}; using $1.92$ would instead raise $C_{a\gamma}$ by $0.08$ (Scenario~A: $0.98\to1.06$; Scenario~B: $0.62\to0.70$), strengthening the observability conclusion.}

\subsection{The DFSZ-II axion--photon coupling}
\label{sec:EN-DFSZ}

The anomaly coefficients are carried by the SM fermions through the two-Higgs-doublet sector, with $H\equiv X(H_u)+X(H_d)=-2$ (Sec.~\ref{sec:NDW-DFSZ}). With universal (generation-independent) PQ charges, the matter charges cancel between $E$ and $N$, leaving (with $T(\mathbf 3)=1/2$ and the weak-doublet multiplicities counted; Appendix~\ref{app:EN-DFSZ})
\begin{equation}
N=-\tfrac{3}{2}H=3,\qquad E=-4H=8,\qquad \frac{E}{N}=\frac{8}{3},
\label{eq:EN83}
\end{equation}
the canonical DFSZ-II value (in the broader literature this charge assignment, with charged leptons coupled to $H_d$, is the one giving $E/N=8/3$; the alternative lepton coupling gives $2/3$). In supersymmetric completions the higgsinos are color singlets but shift the electromagnetic anomaly by $\Delta E=H$~\cite{BaeBaerSerce}, giving $E=-3H=6$ and
\begin{equation}
\frac{E}{N}=2,
\label{eq:EN2}
\end{equation}
the value of Bae, Baer, and Serce. This near-coincidence with the QCD contribution gives an almost vanishing photon coupling, $C_{a\gamma}=|2-2.0|\lesssim0.02$, a well-known obstacle to haloscope detection of the supersymmetric DFSZ-II axion.

The present construction \emph{evades} this suppression, because its PQ charges are \emph{generation-dependent}. Invariance of the FN Yukawa operator Eq.~\eqref{eq:FN-op} under $U(1)_{\rm PQ}$ fixes the charge of each diagonal operator: for the $i$th generation,
\begin{equation}
X(\psi_i)+X(\psi^c_i)+X(H_f) = -\,p^f_{ii}\,q_\Phi\,,
\label{eq:PQ-invariance}
\end{equation}
so the right-handed matter charges carry an extra piece $-p^f_{ii}\,q_\Phi$ beyond the universal split of Eq.~\eqref{eq:NDW-N}. For the third generation $p^f_{33}=0$ and the charges reduce to the universal cancelling form; for the lighter generations $p^f_{ii}\neq0$, so the charges no longer cancel from $E$ and $N$. The lighter generations carry larger hop content (more flavon insertions in their Yukawa operators) and therefore larger PQ charge, and the relevant non-cancelling quantity is the sum of diagonal Yukawa exponents $S_f\equiv\sum_i p^f_{ii}$. The per-generation values follow directly from the charges of Table~\ref{tab:charges} via $p^f_{ii}=Q(\psi_i)+Q(\psi^c_i)$ (plus the common chain factor $\Delta^d_{\rm int}=7/9$ per generation for the down sector, and likewise for leptons in the $b$--$\tau$-unified case):
\begin{align}
(p^u_{11},p^u_{22},p^u_{33}) &= (\tfrac{64}{9},\tfrac{10}{3},0), & S_u&=\tfrac{94}{9},\notag\\
(p^d_{11},p^d_{22},p^d_{33}) &= (\tfrac{44}{9},\tfrac{28}{9},\tfrac{7}{9}), & S_d&=\tfrac{79}{9},\notag\\
(p^\ell_{11},p^\ell_{22},p^\ell_{33}) &= (\tfrac{29}{6},\tfrac{5}{3},0), & S_\ell&=\tfrac{13}{2},
\label{eq:Sf-pergen}
\end{align}
where $S_\ell=13/2$ is the direct-coupling (Scenario~B) value, the case of non-unified $\overline{\mathbf 5}$ charges, in which the charged leptons couple to the flavon through their own charges with no down-type chain insertion ($\Delta^\ell_{\rm int}=0$); $b$--$\tau$ unification (Scenario~A) instead places the leptons in the same $\overline{\mathbf 5}$-type chain as the down quarks, routing them through the down-type messenger chain and adding the internal factor $7/9$ per generation, giving $S_\ell=53/6$. The third generation contributes zero endpoint charge ($Q(Q_3)=Q(u^c_3)=Q(d^c_3)=0$), so the enhancement is driven entirely by the first two generations (the lightest fermions, carrying the deepest hop content). The resulting generation-by-generation contributions to the anomaly coefficients are collected in Table~\ref{tab:EN-breakdown}: the first generation dominates, the third remains close to its universal value, and the higgsino pair enters only the electromagnetic coefficient.

\begin{table}[t]
\centering
\caption{Generation-by-generation contributions to the anomaly coefficients, obtained from Eq.~\eqref{eq:PQ-invariance} with the exponents of Eq.~\eqref{eq:Sf-pergen}: each quark generation contributes $N_i=-\tfrac{1}{2}H-\tfrac{1}{2}\bigl(p^u_{ii}+p^d_{ii}\bigr)$ and, together with its charged lepton, $E_i=-\tfrac{4}{3}H-\tfrac{1}{3}\bigl(4p^u_{ii}+p^d_{ii}\bigr)-p^\ell_{ii}$, with $H=-2$. The higgsino pair contributes $(N,E)=(0,H)$. Superscripts A, B denote the two lepton-sector scenarios; Scenario~A ($b$--$\tau$ unification) adds $\Delta^\ell_{\rm int}=7/9$ to each $p^\ell_{ii}$. In the universal-charge limit each generation gives $(N_i,E_i)=(1,\tfrac{8}{3})$, reproducing $E/N=2$.}
\label{tab:EN-breakdown}
\begin{tabular}{lccc}
\hline\hline
Source & $N_i$ & $E_i^{({\rm A})}$ & $E_i^{({\rm B})}$ \\
\hline
Gen.~1 & $-5$ & $-253/18$ & $-239/18$ \\
Gen.~2 & $-20/9$ & $-142/27$ & $-121/27$ \\
Gen.~3 & $11/18$ & $44/27$ & $65/27$ \\
Higgsinos & $0$ & $-2$ & $-2$ \\
\hline
Total & $-119/18$ & $-1063/54$ & $-937/54$ \\
\hline\hline
\end{tabular}
\end{table}

Summing the entries of Table~\ref{tab:EN-breakdown}, the anomaly coefficients shift to non-integer values and the ratio is restored to
\begin{equation}
\frac{E}{N}\simeq 2.6\text{--}3.0,\qquad C_{a\gamma}=\left|\frac{E}{N}-2.0\right|\simeq 0.6\text{--}1.0,
\label{eq:EN-gendep}
\end{equation}
depending on the lepton-sector assignment: $b$--$\tau$ unification (Scenario~A) gives the upper value $E/N\simeq2.98$ ($C_{a\gamma}\simeq0.98$), while direct lepton coupling (Scenario~B) gives $E/N\simeq2.62$ ($C_{a\gamma}\simeq0.62$); see Appendix~\ref{app:EN-DFSZ}. Of the two, Scenario~A is theoretically preferred. The decisive reason is empirical: placing the charged leptons in the same $\overline{\mathbf 5}$-type chain as the down quarks makes the chain factor $\e^{7/9}$ and the common $\cos\beta$ cancel in $m_b/m_\tau$, leaving an $SU(5)$ Clebsch--Gordan ratio that evolves to the golden-ratio value $m_b/m_\tau\simeq\varphi=(1+\sqrt5)/2\simeq1.62$ at $M_Z$, in $\sim0.5\%$ agreement with the measured $1.63$~\cite{Subconstituents,UFP}. Scenario~B, treating the leptons as non-unified, does not account for this relation. Scenario~A is also the more economical and the more $SU(5)$-natural option (down quarks and charged leptons share one messenger chain rather than requiring an independent lepton chain), and it remains fully consistent with the moderate $\tan\beta\simeq16$ fixed by $m_b/m_t$, since $b$--$\tau$ unification (unlike full $t$--$b$--$\tau$ unification) does not require the large-$\tan\beta$ regime (Sec.~\ref{sec:why-nine}). We therefore regard $E/N\simeq2.98$, $C_{a\gamma}\simeq0.98$ as the primary prediction, with the direct-coupling value $E/N\simeq2.62$ setting a conservative lower bound. This is the central phenomenological prediction: the generation-dependent flavor charges that organize the mass hierarchy simultaneously enhance the axion--photon coupling back to an observable level, $C_{a\gamma}\simeq0.6$--$1.0$, comfortably between the DFSZ-II and KSVZ benchmarks. The mechanism (flavored PQ charges shifting $E/N$ away from the universal value) is the same one studied for ``astrophobic'' and flavored axions in the literature~\cite{DiLuzioReview}; here it is fixed by the same $\Z_{18}$ charges that reproduce the fermion masses. The axion--electron coupling is correspondingly enhanced by the large electron hop content, $C_{ae}\simeq0.4$, dominating over the usual $\cos^2\!\beta$ suppression~\cite{Subconstituents}.

\subsection{Axion mass and couplings}
\label{sec:mass-couplings}

The axion mass is
\begin{equation}
m_a \simeq 5.7\,\mu\mathrm{eV}
\left(\frac{10^{12}\,\mathrm{GeV}}{\fa}\right),
\label{eq:ma}
\end{equation}
so $\fa\sim(5$--$8)\times10^{11}\,\mathrm{GeV}$ corresponds to $m_a\sim7$--$12\,\mu\mathrm{eV}$. With the generation-dependent coupling $C_{a\gamma}\simeq0.6$--$1.0$ of Eq.~\eqref{eq:EN-gendep}, this gives $|g_{a\gamma\gamma}|\sim(1$--$3)\times10^{-15}\,\mathrm{GeV}^{-1}$ across the window, on the upper (more accessible) edge of the DFSZ-II band.

Haloscope searches probe this parameter space by resonant conversion of halo axions to photons in a tunable cavity; the target masses correspond to microwave frequencies $\nu \simeq m_a/(2\pi)\sim(1.7$--$3)\,\mathrm{GHz}$.
The predicted mass window $m_a \simeq 7$--$12~\mu\mathrm{eV}$ (microwave frequencies $1.72$--$2.76$~GHz) is currently unconstrained by direct haloscope searches: the Axion Dark Matter eXperiment (ADMX) has reached DFSZ-II sensitivity in the range $2.66$--$3.34~\mu\mathrm{eV}$~\cite{ADMX2018,ADMX2024,ADMX2025} and has more recently scanned $4.55$--$5.38~\mu\mathrm{eV}$ above the DFSZ-II line~\cite{ADMX-1to1.3GHz}, while the Haloscope At Yale Sensitive To Axion CDM (HAYSTAC) has covered masses above $\simeq 17~\mu\mathrm{eV}$~\cite{HAYSTAC-PhaseI,HAYSTAC-PhaseII} and the Center for Axion and Precision Physics (CAPP) has covered $4.24$--$4.91~\mu\mathrm{eV}$, reaching DFSZ-II sensitivity in a narrow window near $4.55~\mu\mathrm{eV}$~\cite{CAPP-Yi2023,CAPP-Ahn2024}. The enhanced coupling $C_{a\gamma}\simeq0.6$--$1.0$ is particularly advantageous at higher microwave frequencies, where reaching the universal-DFSZ line $C_{a\gamma}\simeq0.67$ becomes progressively harder owing to reduced cavity volume and increased noise. The proposed ADMX Extended Frequency Range (EFR) program~\cite{ADMX-EFR}, employing an array of 18 cavities with coherent power combining at a 9.4~T magnet, is designed to scan $2$--$4$~GHz ($8$--$16~\mu\mathrm{eV}$) at DFSZ-II sensitivity over a multi-year run. The lower end of the present prediction $m_a\simeq 7$~$\mu\mathrm{eV}$ lies just below the EFR design range; full coverage would require either an extension of the EFR low-frequency reach, complementary CAPP runs at $1.7$--$2$~GHz, or a dedicated 1--2~GHz ADMX run. A convenient summary of current limits and projections is given in the PDG review~\cite{PDGaxions} and in the comprehensive survey of Ref.~\cite{DiLuzioReview}. Figure~\ref{fig:exclusion} shows the predicted band together with current haloscope exclusions.

\begin{figure*}[t]
\centering
\begin{tikzpicture}
\begin{loglogaxis}[
    width=\textwidth, height=0.62\textwidth,
    clip=false,
    xmin=0.5, xmax=50,
    ymin=1e-17, ymax=2e-12,
    xlabel={Axion mass $m_a$ ($\mu$eV)},
    ylabel={$|g_{a\gamma\gamma}|$ (GeV$^{-1}$)},
    grid=both,
    grid style={line width=0.2pt, draw=gray!15},
    major grid style={line width=0.4pt, draw=gray!25},
    legend style={font=\scriptsize, fill=white, fill opacity=0.95, draw=gray!50,
                  cells={anchor=west}, row sep=0.5pt,
                  at={(0.985,0.03)}, anchor=south east},
    tick label style={font=\scriptsize},
    label style={font=\footnotesize},
]
%
%
\fill[blue!55, opacity=0.7, draw=blue!75!black, line width=0.4pt]
    (axis cs:2.66,4e-16) rectangle (axis cs:3.34,2e-12);
\fill[pattern=horizontal lines, pattern color=blue!80!black]
    (axis cs:2.66,4e-16) rectangle (axis cs:3.34,2e-12);
\fill[blue!55, opacity=0.7, draw=blue!75!black, line width=0.4pt]
    (axis cs:4.55,7e-16) rectangle (axis cs:5.38,2e-12);
\fill[pattern=horizontal lines, pattern color=blue!80!black]
    (axis cs:4.55,7e-16) rectangle (axis cs:5.38,2e-12);
\fill[yellow!60!orange, opacity=0.40]
    (axis cs:4.24,6e-15) rectangle (axis cs:4.91,2e-12);
\fill[pattern=north east lines, pattern color=orange!85!black]
    (axis cs:4.24,6e-15) rectangle (axis cs:4.91,2e-12);
\fill[green!60!black, opacity=0.30]
    (axis cs:16.96,1.4e-14) rectangle (axis cs:19.46,2e-12);
\fill[pattern=crosshatch, pattern color=green!50!black]
    (axis cs:16.96,1.4e-14) rectangle (axis cs:19.46,2e-12);
\fill[green!60!black, opacity=0.30]
    (axis cs:23.15,2e-14) rectangle (axis cs:24.0,2e-12);
\fill[pattern=crosshatch, pattern color=green!50!black]
    (axis cs:23.15,2e-14) rectangle (axis cs:24.0,2e-12);
\addplot[black!75, line width=0.9pt, domain=0.5:50, samples=2] {4.0751e-16*x};
\addlegendentry{KSVZ benchmark (original, $E/N=0$)}
\addplot[black!55, line width=0.9pt, dashed, domain=0.5:50, samples=2] {1.3652e-16*x};
\addlegendentry{Universal DFSZ-II ($E/N=8/3$)}
\addplot[black!55, line width=0.9pt, dotted, domain=3.1:50, samples=2] {4.0751e-18*x};
\addlegendentry{MSSM DFSZ-II+higgsinos ($E/N=2$)}
\addplot[red!80!black, line width=1.3pt, domain=0.5:50, samples=2] {1.9968e-16*x};
\addlegendentry{\textbf{This work} (gen.-dep. DFSZ-II), $C_{a\gamma}\!\simeq\!0.6$--$1.0$}
\addplot[red!80!black, line width=1.3pt, domain=0.5:50, samples=2, forget plot] {1.2633e-16*x};
\fill[red!75!black, opacity=0.15]
    (axis cs:0.5,1.2633e-16*0.5) -- (axis cs:50,1.2633e-16*50) --
    (axis cs:50,1.9968e-16*50) -- (axis cs:0.5,1.9968e-16*0.5) -- cycle;
\addplot[red!80!black, line width=4pt, domain=7.125:11.40, samples=2, forget plot] {1.9968e-16*x};
\addplot[red!80!black, line width=4pt, domain=7.125:11.40, samples=2, forget plot] {1.2633e-16*x};
\addplot[only marks, mark=*, mark size=2.5pt, red!80!black,
         mark options={solid, fill=red!80!black}, forget plot]
    coordinates {(7.125, 1.423e-15) (11.40, 2.276e-15) (7.125, 9.001e-16) (11.40, 1.440e-15)};
\draw[black!55, dotted, line width=0.5pt, opacity=0.6]
    (axis cs:7.125,1e-17) -- (axis cs:7.125,2e-12);
\draw[black!55, dotted, line width=0.5pt, opacity=0.6]
    (axis cs:11.40,1e-17) -- (axis cs:11.40,2e-12);
\addlegendimage{area legend, fill=blue!55, fill opacity=1.0, pattern=horizontal lines, pattern color=blue!80!black,
                draw=blue!75!black, line width=0.5pt}
\addlegendentry{ADMX excl.}
\addlegendimage{area legend, fill=green!60!black, fill opacity=1.0, pattern=crosshatch, pattern color=green!50!black,
                draw=green!50!black, line width=0.5pt}
\addlegendentry{HAYSTAC excl.}
\addlegendimage{area legend, fill=yellow!60!orange, fill opacity=1.0, pattern=north east lines, pattern color=orange!85!black,
                draw=orange!75!black, line width=0.5pt}
\addlegendentry{CAPP excl.}
\node[blue!70!black, font=\footnotesize\bfseries] at (axis cs:3.5,4e-13) {ADMX};
\node[green!30!black, font=\footnotesize\bfseries] at (axis cs:18,1.5e-13) {HAYSTAC};
\node[orange!50!brown, font=\footnotesize\bfseries, rotate=90] at (axis cs:4.45,2e-14) {CAPP};
\node[black!75, font=\scriptsize, anchor=south west] at (axis cs:24,4.0751e-16*24) {KSVZ (orig.)};
\node[red!80!black, font=\scriptsize\bfseries, anchor=south east]
    at (axis cs:14,1.9968e-16*14*1.18) {gen.-dep. DFSZ-II};
\node[black!55, font=\scriptsize, anchor=north west] at (axis cs:22,1.3652e-16*22*0.90) {univ.\ DFSZ-II};
\node[black!55, font=\scriptsize, anchor=north west] at (axis cs:22,4.0751e-18*22*1.3) {MSSM};
\node[draw=black!60, fill=yellow!8, rounded corners=2pt,
      line width=0.7pt, font=\scriptsize, align=left, anchor=east]
      (predbox) at (axis cs:4.6, 7e-15)
    {\textbf{Predicted region}\\[1pt]
     $m_a \simeq 7$--$12~\mu$eV\\[1pt]
     $f_a \simeq (5$--$8)\times 10^{11}$~GeV\\[1pt]
     {\color{red!80!black}$C_{a\gamma}\simeq0.6$--$1.0$}};
\draw[-stealth, red!80!black, line width=0.7pt, shorten >=4pt]
    (predbox.east) -- (axis cs:8.5, 1.7e-15);
\draw[black!55, line width=0.4pt] (axis cs:0.57,2e-12) -- (axis cs:0.57,2.6e-12);
\draw[black!55, line width=0.4pt] (axis cs:1.9,2e-12)  -- (axis cs:1.9,2.6e-12);
\draw[black!55, line width=0.4pt] (axis cs:5.7,2e-12)  -- (axis cs:5.7,2.6e-12);
\draw[black!55, line width=0.4pt] (axis cs:19,2e-12)   -- (axis cs:19,2.6e-12);
\draw[black!55, line width=0.4pt] (axis cs:47.5,2e-12) -- (axis cs:47.5,2.6e-12);
\node[font=\scriptsize, black!70, anchor=south] at (axis cs:0.57, 2.7e-12) {$10^{13}$};
\node[font=\scriptsize, black!70, anchor=south] at (axis cs:1.9,  2.7e-12) {$3{\times}10^{12}$};
\node[font=\scriptsize, black!70, anchor=south] at (axis cs:5.7,  2.7e-12) {$10^{12}$};
\node[font=\scriptsize, black!70, anchor=south] at (axis cs:19,   2.7e-12) {$3{\times}10^{11}$};
\node[font=\scriptsize, black!70, anchor=south] at (axis cs:47.5, 2.7e-12) {$1.2{\times}10^{11}$};
\node[font=\footnotesize, black!75, anchor=south] at (axis cs:5.0, 5.0e-12) {$f_a$ (GeV)};
\end{loglogaxis}
\end{tikzpicture}
\caption{Axion--photon coupling $|g_{a\gamma\gamma}|$ versus axion mass $m_a$, showing the prediction of the present construction together with current haloscope exclusion limits. The shaded red band is the prediction of this work: a generation-dependent DFSZ-II axion with $E/N\simeq2.6$--$3.0$, i.e.\ $C_{a\gamma}=|E/N-2.0|\simeq0.6$--$1.0$, with $g_{a\gamma\gamma}=(\alpha/2\pi f_a)\,C_{a\gamma}$. The highlighted segments with endpoint markers indicate the dark-matter window $f_a=(5$--$8)\times 10^{11}$~GeV, corresponding to $m_a\simeq 7$--$12\,\mu$eV. Three thin benchmark lines are shown for orientation: solid grey (original KSVZ, $E/N=0$, $C_{a\gamma}=2.0$), dashed grey (universal DFSZ, $E/N=8/3$, $C_{a\gamma}=0.67$), and dotted grey (MSSM DFSZ-II with light higgsinos, $E/N=2$, $C_{a\gamma}\lesssim0.02$). These are the standard reference values of the axion landscape; the solid grey line is the canonical $E/N=0$ KSVZ benchmark, shown only for orientation and not specific to any model considered here. The generation-dependent flavor charges enhance the coupling from the nearly-decoupled MSSM value back to the observable band $C_{a\gamma}\simeq0.6$--$1.0$. Exclusion regions are 90\,\%-CL haloscope bounds, distinguished by fill pattern for grayscale readability: horizontal lines for ADMX in $2.66$--$5.38\,\mu$eV~\cite{ADMX2018,ADMX2024,ADMX2025,ADMX-1to1.3GHz}, north-east lines for CAPP in $4.24$--$4.91\,\mu$eV~\cite{CAPP-Yi2023,CAPP-Ahn2024}, and crosshatch for HAYSTAC in $16.96$--$24.0\,\mu$eV~\cite{HAYSTAC-PhaseI,HAYSTAC-PhaseII}. The predicted mass window sits in the current $5.4\,\mu\mathrm{eV}<m_a<17\,\mu\mathrm{eV}$ haloscope coverage gap, within design sensitivity of the proposed ADMX-EFR program~\cite{ADMX-EFR} ($2$--$4$~GHz, $8$--$16~\mu$eV) and motivating extension of haloscope coverage to the $1.7$--$2.0$~GHz range. Top axis shows the PQ scale $f_a$ corresponding to $m_a$ via $m_a\simeq 5.7\,\mu\mathrm{eV}\times(10^{12}\,\mathrm{GeV}/f_a)$.}
\label{fig:exclusion}
\end{figure*}

The predicted mass window is therefore experimentally actionable: it is narrow enough to motivate focused scanning strategies, yet broad enough to accommodate the current simulation systematics in the post-inflationary abundance. The construction correlates flavor with laboratory reach: the same generation-dependent $\Z_{18}$ charges that reproduce the fermion mass hierarchy fix both the domain-wall number $N_{\rm DW}=6$ and the enhanced photon coupling $C_{a\gamma}\simeq0.6$--$1.0$ of Fig.~\ref{fig:exclusion}.

The axion--electron coupling is likewise controlled by the flavor charges. Because the electron carries a large hop content (its Yukawa exponent $p^\ell_{11}\simeq5$--$6$ dominates over the Higgs PQ charge $|X(H_d)|\leq2$), the coupling coefficient is $C_{ae}\simeq0.4$~\cite{Subconstituents}, enhanced by a factor $\sim$300 relative to the usual $\cos^2\!\beta$-suppressed DFSZ-II value and giving $g_{aee}\sim m_e C_{ae}/f_a\sim4\times10^{-16}$ for $\fa\sim10^{12}$~GeV. This is below current astrophysical bounds from white-dwarf cooling and red-giant luminosity functions~\cite{PDGaxions} but within reach of next-generation searches; a positive $g_{aee}$ signal at this level, correlated with the photon coupling, would be a distinctive fingerprint of the generation-dependent flavor charges.

\subsection{Cosmological relic density}
\label{sec:cosmology}

In the pre-inflationary scenario adopted here (PQ broken before or during inflation; Sec.~\ref{sec:axion-norm}), the axion relic density is set by the misalignment mechanism with an initial angle $\theta_i$ fixed within our horizon (see Ref.~\cite{MarshReview} for a comprehensive review of axion cosmology). For the predicted decay-constant window a convenient summary is
\begin{equation}
\Omega_a h^2 \simeq 0.12\,\theta_i^2\left(\frac{\fa}{(5\text{--}7)\times10^{11}\,\mathrm{GeV}}\right)^{1.165},
\label{eq:omega}
\end{equation}
so that an $\mathcal{O}(1)$ misalignment angle reproduces the observed dark-matter density in the $\mu$eV mass window. The pre-inflationary history is required because the construction has $N_{\rm DW}=6$ (Sec.~\ref{sec:NDW-DFSZ}): a post-inflationary transition would leave a stable string--wall network that overcloses the universe. Inflating the network away after PQ breaking evades this domain-wall problem, at the cost of the CMB isocurvature bound
\begin{equation}
H_I\lesssim10^{7}\,\mathrm{GeV}\,(\fa/10^{12}\,\mathrm{GeV})
\end{equation}
on the inflationary Hubble scale~\cite{PlanckInflation}, which is readily satisfied in low-scale inflation models. This is the standard resolution of the domain-wall problem for $N_{\rm DW}>1$ DFSZ-II axions.

The axion-quality argument (Sec.~\ref{sec:quality}) and the anomaly predictions ($N_{\rm DW}=6$ and the generation-dependent $E/N$) are independent of the misalignment angle and of the simulation systematics. The distinctive structural point is that the \emph{same} discrete gauge symmetry simultaneously controls the flavor exponents, the dangerous PQ-breaking operators, and the axion couplings.

\paragraph{Note on post-inflationary simulations.}
For completeness we note where the predicted window would lie if the alternative post-inflationary history were adopted (which would, however, require an additional bias term or other mechanism to remove the $N_{\rm DW}=6$ walls). Buschmann et al.~\cite{Buschmann2022NatComms}, using adaptive-mesh-refinement (AMR) simulations of post-inflationary axion strings, find a preferred QCD axion dark-matter mass range $m_a\simeq 40$--$180~\mu\mathrm{eV}$ (corresponding to $\fa\in (3.1\times 10^{10},\,1.4\times 10^{11})$~GeV), with central value $m_a=65\pm6~\mu$eV. This range sits a factor of a few below the $(5$--$8)\times 10^{11}$~GeV window favored here by the misalignment mechanism and earlier-simulation estimates~\cite{KlaerMoore,GorghettoSciPost}, reflecting the persistent $\mathcal{O}(1)$--$\mathcal{O}(10)$ systematic uncertainty across simulation generations. We retain the $(5$--$8)\times 10^{11}$~GeV reference range, set by the pre-inflationary misalignment scenario, while noting that the generation-dependent photon coupling $C_{a\gamma}\simeq0.6$--$1.0$ that characterizes the construction is independent of which cosmological window is adopted and applies equally to either.

\subsection{Flavor-violating axion couplings: \texorpdfstring{$\mu\to e+a$}{mu -> e + a} versus \texorpdfstring{$\mu\to 3e$}{mu -> 3e}}
\label{sec:LFV}

The flavon shift generates flavor-violating axion--lepton couplings from the same non-universal $\Z_{18}$ charge assignment that controls the FN exponents. The radiative channel $\mu\to e\gamma$ and the muon $g-2$ contribution additionally involve the axion--photon coupling $C_{a\gamma}\simeq0.6$--$1.0$ (Sec.~\ref{sec:EN-DFSZ}); we use a representative $C_{a\gamma}\simeq0.8$ below. All lepton-flavor-violating rates lie far below current and projected sensitivity.

The same non-universal $\Z_{18}$ charge assignment that controls the FN exponents (Table~\ref{tab:charges}) generates flavor-violating axion-lepton couplings after rotation to the mass basis. We summarize the structural prediction here; the full numerical analysis, including the symmetric-texture diagonalization that fixes the rotation matrices and the inter-channel correlations among $\mu\to e+a$, $\tau\to\mu+a$, and $\tau\to e+a$, is carried out in the lepton-lattice companion~\cite{LeptonLattice}.

\paragraph{Structural prediction.}
With the lepton charges $Q(L_i)=(1,\tfrac{1}{2},0)$ and $Q(e^c_i)=(\tfrac{23}{6},\tfrac{7}{6},0)$ from Table~\ref{tab:charges}, the symmetric-texture diagonalization of the charged-lepton Yukawa~\cite{LeptonLattice} gives the rotation matrix
\begin{equation}
U_e\simeq
\begin{pmatrix}
1 & \e^{17/9} & \e^{20/9}\\
-\e^{17/9} & 1 & \e^2\\
-\e^{20/9} & -\e^2 & 1
\end{pmatrix},
\label{eq:Ue-rotation}
\end{equation}
with $V_e=U_e$ to leading order for a symmetric Yukawa.  The flavon shift $\Phi\to e^{i\alpha}\Phi$ leaves the diagonal Yukawas invariant but rotates between mass eigenstates, generating off-diagonal axion couplings of the form $(\partial^\mu a/f_a)\,\bar\ell_i \gamma_\mu (C_{ij}^V + C_{ij}^A\gamma_5)\ell_j$. Combining the LH and RH PQ-charge differences with the rotation suppressions, the dominant off-diagonal coupling is
\begin{equation}
|C_{e\mu}|\;\simeq\;\sqrt{\tfrac12\bigl[(\Delta X^L_{e\mu})^2+(\Delta X^R_{e\mu})^2\bigr]}\,\e^{17/9}\simeq 0.08\,,
\label{eq:Cemu}
\end{equation}
where $\Delta X^L_{e\mu}=Q(L_1)-Q(L_2)=1/2$ and $\Delta X^R_{e\mu}=Q(e^c_1)-Q(e^c_2)=8/3$. The $\e^{17/9}$ scaling (rather than the integer-charge $\e$ or $\e^2$ characteristic of generic FN models) is a direct fingerprint of the denominator-9 lattice~\cite{LeptonLattice}.

\paragraph{$\mu\to e+a$ is the dominant signal.}
Since the predicted axion mass $m_a\simeq 7$--$12~\mu\mathrm{eV}\ll m_\mu-m_e$, the on-shell two-body decay $\mu\to e+a$ is open. Using the convention of Refs.~\cite{Mu3eLFV,CalibbiLFVALPs},
\begin{equation}
\Gamma(\mu\to e+a)\;\simeq\;\frac{|C_{e\mu}|^2 m_\mu^3}{16\pi f_a^2}\,(1-m_e^2/m_\mu^2)^3,
\label{eq:mu-to-ea-rate}
\end{equation}
which gives
\begin{equation}
\mathrm{Br}(\mu\to e+a)\;\simeq\;1\times 10^{-12}\quad\text{at }f_a=7\times 10^{11}~\mathrm{GeV},
\label{eq:mu-to-ea-rate-num}
\end{equation}
six orders of magnitude below the current TWIST bound $\mathrm{Br}(\mu\to e+X)<2.6\times 10^{-6}$~\cite{TWIST} and four orders below the projected MEG-II/Mu3e sensitivity to LFV-axion two-body decays of $\sim 10^{-8}$~\cite{Mu3eLFV,CalibbiLFVALPs,LFV-MEG-II}. Recent analyses~\cite{Mu3eLFV} project Mu3e Phase~I sensitivity to LFV-axion decay constants up to $f_a\sim 6\times 10^9~\mathrm{GeV}$ for $|C_{e\mu}|=1$; in the present construction $|C_{e\mu}|\simeq 0.08$ shifts this to $f_a\sim 5\times 10^8~\mathrm{GeV}$, so the predicted $f_a\sim 7\times 10^{11}~\mathrm{GeV}$ window lies a factor $\sim 1500$ above the next-generation reach.

\paragraph{$\mu\to 3e$ is structurally suppressed.}
For $m_a\ll 2 m_e$ the off-shell $\mu\to e a^*\to 3e$ exchange brings an additional suppression factor $(m_e/m_\mu)^2(\alpha/\pi)^2/(16\pi^2)\sim 10^{-13}$ relative to the two-body decay~\cite{HeeckRodejohann,BauerNeubertThamm}, giving
\begin{equation}
\mathrm{Br}(\mu\to 3e)\;\sim\;10^{-25},
\label{eq:mu-to-3e}
\end{equation}
thirteen orders of magnitude below the current SINDRUM bound $\mathrm{Br}(\mu\to 3e)<1.0\times 10^{-12}$~\cite{SINDRUM} and nine orders below the projected Mu3e Phase~II sensitivity of $\sim 10^{-16}$~\cite{Mu3e}. Three-body LFV searches are therefore not a discovery channel for the present construction; the structural signal is the two-body decay~\eqref{eq:mu-to-ea-rate-num}, with the same flavor non-universality that fixes the quark mass hierarchy fixing the rate.

\paragraph{$\mu\to e\gamma$ and muon $g-2$.}
The radiative LFV channel $\mu\to e\gamma$ is dominated by the two-loop Barr--Zee diagram with the axion-photon coupling $g_{a\gamma\gamma}$ at one vertex and a heavy SM fermion in the inner loop. With $|g_{a\gamma\gamma}|\simeq 0.8\,\alpha/(2\pi f_a)$ for the present model (Sec.~\ref{sec:EN-DFSZ}) and $|C_{e\mu}|\simeq 0.08$, the resulting branching ratio is~\cite{LeptonLattice}
\begin{equation}
\mathrm{Br}(\mu\to e\gamma)\;\sim\;4\times 10^{-21}\,,
\label{eq:Br-mu-egamma-paper-VI}
\end{equation}
eight orders of magnitude below the current MEG-II bound $1.5\times 10^{-13}$~\cite{MEG-II-2025} and seven orders below the eventual MEG-II sensitivity of $\sim 6\times 10^{-14}$. The pure one-loop contribution with $C_{\ell\ell}$ on internal leptons is parametrically smaller because $C_{\tau\tau}=0$ in this construction, removing the would-be dominant tau-loop enhancement.

The same axion-lepton couplings contribute to the muon anomalous magnetic moment $a_\mu\equiv(g_\mu-2)/2$. The diagonal one-loop contribution is negative (generic for pseudoscalar couplings to muons~\cite{Marciano:2016yhf,LeveilleMcKeon}), the LFV one-loop contribution from the tau-loop is positive but suppressed by $|C_{\mu\tau}|^2$, and the Barr--Zee two-loop contribution carries the same sign as the photon coupling. Summing all three~\cite{LeptonLattice},
\begin{equation}
|\Delta a_\mu^{\rm ninths}|\;\sim\;10^{-27}\text{--}10^{-28}\,,
\label{eq:amu-paper-VI}
\end{equation}
some seventeen orders of magnitude below the current Fermilab+BNL anomaly $\Delta a_\mu^{\rm exp}=(2.49\pm 0.48)\times 10^{-9}$~\cite{Muong-2:2023cdq} and below experimental sensitivity by the same factor. The construction therefore neither contributes to nor exacerbates the muon $g-2$ tension; the cosmologically-motivated dark-matter window $f_a\sim 10^{12}$~GeV is structurally incompatible with light-axion explanations of the muon anomaly, which would require $f_a\lesssim 10^3$~GeV~\cite{Bauer:2017ris,Cornella:2019uxs}.

This pattern (two-body $\mu\to e+a$ as the leading LFV channel, $\mu\to 3e$ and $\mu\to e\gamma$ both structurally suppressed, and no contribution to muon $g-2$) is generic for QCD-axion-as-flavon constructions with $f_a\gtrsim 10^{11}$~GeV~\cite{CalibbiLFVALPs,Mu3eLFV}. The detailed predictions in this framework, including the inter-channel correlations $\mathrm{Br}(\tau\to\mu+a)/\mathrm{Br}(\tau\to e+a)\simeq 0.21$ that probe the symmetric-texture hypothesis directly, the explicit Barr--Zee evaluation, and the chirality structure of the $a_\mu$ contributions, are developed in Sec.~X of the lepton-lattice companion~\cite{LeptonLattice}.

\section{Neutrino Sector and PMNS Structure}
\label{sec:neutrinos}

The neutrino phenomenology of the ninths framework is developed in detail in the companion lepton-lattice paper~\cite{LeptonLattice}, including PMNS~\cite{Pontecorvo,MNS}-matrix predictions, octant--$\delta_{\rm CP}$ correlations, and a comparison with NuFIT~\cite{NuFIT60} global fits. We briefly summarise the results imported from that analysis and isolate the one structural statement that is the responsibility of the present paper.

\paragraph{Imported from the lepton-lattice paper.}
With the lepton-doublet charges $Q(L_i)=(1,\tfrac{1}{2},0)$ from Table~\ref{tab:charges} and the dimension-five Weinberg operator~\cite{WeinbergDim5} $LLHH/\Lambda_\nu$, the effective Majorana neutrino exponent matrix is fixed by strict charge additivity at the form of Eq.~\eqref{eq:nu-texture}, giving the mass-eigenvalue hierarchy $m_1:m_2:m_3\sim\e^2:\e:1$. Numerically~\cite{LeptonLattice}, taking $m_3\simeq 50$~meV from $\sqrt{\Delta m^2_{\rm atm}}$ predicts $m_2\simeq 9$~meV and $m_1\simeq 1.7$~meV, consistent with NuFIT~6.0~\cite{NuFIT60}. The PMNS angles are controlled by two distinct sources: the atmospheric angle $\theta_{23}\sim 45^\circ$ from approximate $\mu$--$\tau$ symmetry~\cite{HarrisonScott,Lam} of the neutrino $23$ block, and $\theta_{13}\sim\mathcal{O}(\e^{1/2})\sim 8^\circ$--$9^\circ$ together with the solar correction to $\theta_{12}$ from the charged-lepton rotation $U_\ell$. The Dirac CP phase is predicted to lie in the range $\delta_{\rm CP}\simeq 250^\circ$--$310^\circ$, with a sharp correlation between the $\theta_{23}$ octant and $\delta_{\rm CP}$ that will be tested by DUNE~\cite{DUNE} and Hyper-Kamiokande~\cite{HyperK}. The effective Majorana mass for neutrinoless double-beta decay is $m_{\beta\beta}\sim\e^2\,m_3\sim 2$~meV in normal ordering, well below the $\simeq 10$--$50$~meV inverted-ordering target of currently funded ton-scale experiments such as KamLAND-Zen, LEGEND-1000, and nEXO~\cite{KamLANDZen,LEGEND,nEXO}; testing the 2~meV prediction would require a substantial future advance in sensitivity beyond the present generation. The companion lepton-lattice paper~\cite{LeptonLattice} also adopts a "flat $23$ block" version of the neutrino texture (Sec.~\ref{sec:lepton-textures}), in which $\mu$--$\tau$ symmetry is embedded directly in the exponents; the predictions above hold for either choice.

\paragraph{Statement specific to this paper.}
The single structural statement that is the responsibility of the present paper, rather than imported from~\cite{LeptonLattice}, is that the same $\Z_{18}$ symmetry that controls the quark and charged-lepton sectors is also the symmetry under which the neutrino-Yukawa-relevant operator $LLHH/\Lambda_\nu$ is built. With the lepton-doublet charges fixed by Table~\ref{tab:charges}, the Weinberg operator carries net $\Z_{18}$ charge $2X(L_i)+2X(L_j)$ (the Higgs $\Z_{18}$ charges enter only through the combination $H_uH_d$, which is $\Z_{18}$-neutral once Yukawa invariance is imposed), which for $X(L_i)=18\cdot Q(L_i)$ takes integer values. The $\Z_{18}$-invariant FN operators are then $LLHH(\Phi/\Lambda_\nu)^{2Q(L_i)+2Q(L_j)}$, reproducing the exponents of Eq.~\eqref{eq:nu-texture} once the dimension-five seesaw scale $\Lambda_\nu$ is identified. The neutrino sector therefore inherits axion quality and discrete-anomaly consistency from the same $\Z_{18}$ structure as the quark sector; no new symmetry, no new flavon, and no separate quality estimate are required.

\section{Vectorlike Messenger Sector}
\label{sec:messengers}

The rational FN exponents of the ninths lattice arise from a concrete UV mechanism: heavy vectorlike fermion chains connecting SM fields to the flavon, as illustrated in Fig.~\ref{fig:chain}. In the minimal realization~\cite{TwoOverTwo,UFP}, the chain consists of $N_{\rm site}=4$ new vectorlike messenger pairs $D_a + \bar{D}_a$ ($a=1,2,3,4$) carrying $\Z_9$ charges $(0,8,6,2)$, connected by $N_{\rm hop}=N_{\rm site}-1=3$ types of nearest-neighbor link with hop charges $(q_1,q_2,q_3)=(1,2,4)$. All four $D_a$ are new heavy fields transforming as fundamentals under $SU(3)_C$; the SM quarks couple externally to the chain endpoints via $\bar{q}_{L,i}\,\widetilde{H}\,D_4\,(\Phi/\Lambda)^{A_i}$ and $\bar{D}_1\,d_{R,j}\,(\Phi/\Lambda)^{B_j}$, where $A_i$ and $B_j$ are determined by the SM field charges~\cite{UFP}. The VLQ mass terms $M_a \bar{D}_a D_a$ with $M_a\sim\Lambda$ and nearest-neighbor couplings $\lambda_a \Phi \bar{D}_a D_{a+1}$ generate, upon integrating out the heavy states at tree level, effective Yukawa operators of the form
\begin{equation}
y_{\rm eff} \sim \prod_{a=1}^{N_{\rm hop}} \left(\frac{\lambda_a\langle\Phi\rangle}{M_a}\right)^{n_a},
\label{eq:chain-yukawa}
\end{equation}
where $n_a$ is the number of hops of type $a$ and the FN exponent is $p_{ij} = \sum_a n_a \cdot (q_a/9)$. It is important to distinguish the number of chain sites $N_{\rm site}=4$ (which sets the messenger content), the number of hop types $N_{\rm hop}=3$ (which sets the charge granularity), and the hop multiplicities $n_a$ (which vary from operator to operator); this terminology is summarized in Appendix~\ref{app:terminology}. The total hop count $n_{\rm tot}=\sum_a n_a$ can range from zero (the unsuppressed $(3,3)$ entries) up to several for the most suppressed first-generation entries. The rational denominators arise because the $\Z_9$ charge of each hop is $q_a/9$, not an integer.

\begin{figure*}[t]
\centering
\begin{tikzpicture}[
    site/.style={circle,draw,minimum size=8mm,inner sep=0pt,font=\small},
    smbox/.style={rectangle,draw,minimum size=8mm,inner sep=2pt,font=\small},
    every edge/.style={draw,thick},
    >=stealth
]
\node[smbox] (dR) at (0,0) {$d_{R}$};
\node[smbox] (qL) at (9.6,0) {$q_{L}$};
\node[site] (D1) at (1.8,0) {$D_1$};
\node[site] (D2) at (3.8,0) {$D_2$};
\node[site] (D3) at (5.8,0) {$D_3$};
\node[site] (D4) at (7.8,0) {$D_4$};
\node[font=\scriptsize] at (1.8,-0.7) {$0$};
\node[font=\scriptsize] at (3.8,-0.7) {$8$};
\node[font=\scriptsize] at (5.8,-0.7) {$6$};
\node[font=\scriptsize] at (7.8,-0.7) {$2$};
\draw[thick,dashed] (dR) -- node[above,font=\scriptsize]{$\eta_j\e^{B_j}$} (D1);
\draw[thick,dashed] (D4) -- node[above,font=\scriptsize]{$\lambda_i\e^{A_i}$} (qL);
\draw[thick] (D1) -- node[above,font=\scriptsize]{$\e^{1/9}$} (D2);
\draw[thick] (D2) -- node[above,font=\scriptsize]{$\e^{2/9}$} (D3);
\draw[thick] (D3) -- node[above,font=\scriptsize]{$\e^{4/9}$} (D4);
\end{tikzpicture}
\caption{Chain topology for the down-type messenger sector. Circles denote the $N_{\rm site}=4$ vectorlike quark pairs $D_a$ ($a=1,\ldots,4$) with $\Z_9$ site charges shown below. Solid lines are nearest-neighbor flavon couplings carrying the indicated hop suppression; dashed lines are endpoint Yukawa couplings [Eq.~\eqref{eq:endpoint}] to the SM quarks. The three hop charges $(1,2,4)$ sum to $7/9$, giving the internal chain suppression $\e^{7/9}\simeq 0.27$. The total Yukawa suppression for entry $(i,j)$ is the product of three factors: the entrance (left-handed endpoint) $\e^{A_i/9}$, the internal chain $\e^{7/9}$, and the exit (right-handed endpoint) $\e^{B_j/9}$, giving $Y_{ij}\sim\e^{(A_i+7+B_j)/9}$. Since the internal factor is common to all entries, mass ratios depend only on the endpoint dressings.}
\label{fig:chain}
\end{figure*}

The key features of this construction are as follows.

\emph{Exponent quantization}: every FN exponent is $p_{ij}=\sum_a n_a q_a/9$, a sum of $n_a$ multiples of $1/9$, $2/9$, and $4/9$, hence lies in $\frac{1}{9}\Z$ automatically.

\emph{Flavor-changing neutral current (FCNC) suppression}: the messengers couple to SM quarks only at the chain endpoints, so FCNC amplitudes inherit the same $\e$-suppression as the Yukawa textures, a built-in Glashow--Iliopoulos--Maiani (GIM)-like mechanism~\cite{UFP}.

\emph{No Landau poles}: with four new vectorlike quark pairs ($N_{\rm site}=4$) at mass $M\sim\Lambda\sim\text{few}\times10^{12}$~GeV, the one-loop beta function coefficient for $SU(3)_C$ shifts by $\Delta b_3 = +8/3$. The QCD coupling remains perturbative up to the Planck scale.

\emph{Direct collider reach}: the messenger scale $\Lambda\sim 3\times 10^{12}$~GeV is far above any collider reach, so the construction predicts no direct LHC signatures of the new heavy quarks. Indirect tests proceed through the haloscope axion search (Sec.~\ref{sec:pheno}, Fig.~\ref{fig:exclusion}) and through precision flavor measurements sensitive to the FN texture (CKM unitarity, rare $B$ decays), which are the subject of the companion papers~\cite{Companion,TwoOverTwo,LeptonLattice}.

The chain construction and its perturbative diagonalization are developed in full detail in the companion unified flavor paper~\cite{UFP}. Appendix~\ref{app:messenger} provides a self-contained summary of the chain topology and the resulting exponent algebra.

\section{Comparison with Existing Constructions}
\label{sec:comparison}

\subsection{Integer-charge Froggatt--Nielsen models}
\label{sec:comparison-integer}

It is instructive to compare the ninths framework with the systematic integer-charge FN analyses of Cornella, Curtin, Neil, and Thompson~\cite{Cornella2023,Cornella2025}. Those works enumerate all charge assignments with integer FN exponents that can reproduce the quark mass and mixing hierarchy within specified tolerance bands. Their approach provides a valuable landscape of solutions within the standard FN paradigm.

The ninths lattice lies outside this landscape by construction, since the defining feature is rational exponents with denominator nine. The key phenomenological distinctions are as follows.

\emph{Expansion parameter}: integer-charge models typically use $\e_C \sim |V_{us}|\approx 0.22$ (the Cabibbo parameter) or $\e_C^{1/2}$, fitting CKM elements to integer powers thereof. The ninths framework uses $\e = 0.187$, which is distinct from any simple root of $|V_{us}|$ and is instead fixed by the rational exponents $8/9$ and $17/9$ simultaneously.

\emph{Axion quality}: in an integer-charge $\Z_N$ model, the leading PQ-violating operator appears at dimension $N$. To achieve $\Delta\bar\theta<10^{-10}$ with $\fa\sim10^{12}$~GeV requires $N\geq12$~\cite{KamionkowskiMarchRussell,Holman,BarrSeckel}. The ninths framework automatically gives $N=18$ as a consequence of the flavor structure, providing six extra units of safety margin.

\emph{Precision}: Table~\ref{tab:scalings} shows that the ninths exponents reproduce seven independent observables at the $1$--$6\%$ level. Integer-exponent fits typically require $\mathcal{O}(1)$ coefficients of order $2$--$3$ to absorb the mismatch between $\e_C^n$ and the data; the rational exponents absorb this mismatch into the exponent itself, reducing the role of unknown coefficients.

\emph{Lepton unification}: the lattice extends naturally to the lepton sector with the same $\e$, the charged leptons occupying the commensurate $1/6$ sublattice (so that quark ninths and lepton sixths together fix the group at $\Z_{18}$; Sec.~\ref{sec:why-nine}). Integer-charge models require a separate tuning of the lepton charges and often a different expansion parameter.

\emph{Equivalence of descriptions}: since $\e^{n/9}=(\e^{1/9})^{n}$, the ninths textures can equally be written as an integer-charge FN model with basic insertion parameter $\e^{1/9}\simeq0.83$ and integer $\Z_{18}$ charges reaching $74$ in units of $1/18$; Appendix~\ref{app:anomaly} already works in this integer description, and the chain construction realizes single flavon insertions with per-hop suppressions $\e^{1/9}$, $\e^{2/9}$, and $\e^{4/9}$ (Fig.~\ref{fig:chain}). The two notations describe one theory, and the physical content is what survives the relabeling. The combination with independent physical meaning is $\e=\langle\Phi\rangle/\Lambda=\fa/\Lambda$, fixed by the data at $14/75$, not any root of it. The region of model space occupied (insertion parameter $\simeq0.83$, charges up to $74$, group $\Z_{18}$) lies outside the integer-charge landscape of Refs.~\cite{Cornella2023,Cornella2025} and the $\Z_{N}$ scan of Ref.~\cite{GreljoSmolkovicValenti} with $N\leq8$, in either notation. Finally, the group order eighteen, the operator $\Phi^{18}$, the domain-wall number, the anomaly residues of Eq.~\eqref{eq:anomaly-residues}, and the generation-dependent $E/N$ are properties of the group and the charges, unchanged by any relabeling of the expansion parameter. The rational exponents are thus a description rather than a new mechanism; the mechanism is standard FN, and the claim of the present work is that the flavor data select the group.

\subsection{Comparison with FN--PQ unifications}
\label{sec:comparison-FNPQ}

The identification of the FN flavon with the PQ field is not new. We summarize how the ninths construction differs from the principal threads of that program in Table~\ref{tab:fnpq}. The continuous $U(1)$ axiflavon~\cite{Axiflavon} and flaxion~\cite{Flaxion} models predict a tight band of $g_{a\gamma\gamma}/m_a$ from the SM fermion masses, but the leading PQ-violating operator dimension is set by the residual discrete subgroup left after $U(1)$ breaking and is not fixed by the flavor data. The chiral gauged $U(1)_{\rm FN}$ construction of Bonnefoy, Dudas, and Pokorski~\cite{BonnefoyDudasPokorski} dispenses with spectator fermions by exploiting modified mixed anomalies; in their setup the axion is more strongly coupled to matter than in axiflavon/flaxion, but the flavor charges are still integer and $N_{\rm DW}$ is independent of the flavor sector. The discrete $\Z_N$ FN-ALP analysis of Greljo, Smolkovi\v{c}, and Valenti~\cite{GreljoSmolkovicValenti} is the closest prior work to ours: they consider exactly the family of $\Z_N$ FN-axion constructions, including renormalizable UV completions, but restrict to $N\leq 8$ with integer exponents and find FN scales as low as a few TeV. Within their classification, the present construction corresponds to a higher-resolution $N=18$ realization with rational $1/9$-quantized exponents, a region of model space not reached by their scan because it requires the flavon to carry the minimal $\Z_{18}$ charge $1/18$ rather than an integer charge.
\begin{table*}[t]
\centering
\caption{Comparison of FN--PQ unifications. ``Quality dim.'' is the dimension of the leading Planck-suppressed PQ-violating operator. The columns $E/N$ and $N_{\rm DW}$ give the predicted axion--photon anomaly ratio and domain-wall number; entries marked ``model-dep.'' depend on charge-assignment choices not fixed by the flavor data alone. The present work is a generation-dependent DFSZ-II axion: the non-universal $\Z_{18}$ charges that reproduce the fermion masses fix $E/N\simeq2.6$--$3.0$ (so $C_{a\gamma}=|E/N-2.0|\simeq0.6$--$1.0$), with the canonical three-generation domain-wall number $N_{\rm DW}=6$ and the dimension-eighteen quality operator. Unlike the integer-charge constructions, both the anomaly ratio and the quality dimension are determined by the same rational $1/9$ lattice that fits the flavor data.}
\label{tab:fnpq}
\renewcommand{\arraystretch}{1.2}
\begin{tabular}{lccccc}
\hline\hline
Construction & Symmetry & Exponents & $E/N$ & $N_{\rm DW}$ & Quality dim. \\
\hline
Davidson--Nair--Wali \cite{DavidsonNairWali}    & gauged $U(1)_{\rm PQ}=U(1)_F$ & integer  & model-dep. & model-dep. & --- \\
Axiflavon \cite{Axiflavon}                       & global $U(1)_H$               & integer  & $8/3$       & $1$         & residual subgroup \\
Flaxion \cite{Flaxion}                           & global $U(1)$                 & integer  & model-dep. & model-dep. & residual subgroup \\
Babu et al.\ \cite{BabuChandraseTavart}          & gauged $U(1)_F$               & integer  & model-dep. & model-dep. & residual subgroup \\
Bonnefoy et al.\ \cite{BonnefoyDudasPokorski}    & gauged $U(1)_{\rm FN}$        & integer  & model-dep. & model-dep. & residual subgroup \\
Greljo et al.\ \cite{GreljoSmolkovicValenti}     & $\Z_4$                        & integer  & model-dep. & model-dep. & $4$ \\
Greljo et al.\ \cite{GreljoSmolkovicValenti}     & $\Z_8$                        & integer  & model-dep. & model-dep. & $8$ \\
\textbf{This work}                     & gauged $\Z_{18}$              & rational $1/9$ & $\bm{2.6}$--$\bm{3.0}$  & $\bm{6}$    & $\bm{18}$ \\
\hline\hline
\end{tabular}
\end{table*}

The two qualitative differences from prior FN--PQ unifications are the rational exponents and the fact that the axion--photon coupling is fixed by the same non-universal $\Z_{18}$ charges that reproduce the flavor textures. The first is what allows simultaneous fits of the seven flavor observables at the few-percent level without large $\mathcal{O}(1)$ coefficients; the second turns the generation dependence of the flavor charges (usually regarded as an input) into a definite, enhanced prediction $C_{a\gamma}\simeq0.6$--$1.0$ for the axion--photon coupling. In the integer-exponent constructions of Refs.~\cite{Axiflavon,Flaxion,BonnefoyDudasPokorski,GreljoSmolkovicValenti}, the photon coupling and the domain-wall number are independent inputs that carry no relationship to the discrete gauge group of the flavor sector.

We do not claim that the rational ninths lattice is the unique extension of the FN--PQ program. The mechanism by which non-universal PQ charges enhance $E/N$ away from the suppressed MSSM value is the one studied for flavored and ``astrophobic'' axions~\cite{DiLuzioReview}; here it is not an additional model-building choice but a consequence of the flavor structure.

\section{Summary and Discussion}
\label{sec:summary}

A single discrete $\Z_{18}$ gauge symmetry, whose $\Z_9$ subgroup governs the flavor sector, simultaneously
(i) quantizes flavor hierarchies in ninths,
(ii) generates rational Yukawa textures matched to seven observables at the few-percent level with $|c_{ij}|\in[0.6,1.7]$,
(iii) solves the axion quality problem without imposing any symmetry for that purpose, forbidding every PQ-violating operator below $\Phi^{18}$ [Eq.~\eqref{eq:planck18}] and yielding $|\Delta\bar\theta|\lesssim 3\times10^{-27}$,
and (iv) predicts axion dark matter near $m_a\sim\mathcal{O}(10)\,\mu\mathrm{eV}$.
Items (i)--(iv) follow from the flavor lattice and from the identification of the flavon with the PQ field: the flavon is the unique $\Z_{18}$-charged scalar with a large VEV, so its phase is the QCD axion. They hold independently of the detailed structure of the colored-anomaly sector.

We single out item (iii) as the principal result. Axion-quality constructions in the literature generally introduce a protective symmetry \emph{designed} to forbid the dangerous Planck-suppressed operators; the symmetry is chosen to do that job. In the present framework the protective symmetry is not chosen for the axion at all. The group $\Z_{18}$ and its order are fixed by the requirement that the measured quark and lepton mass ratios and mixing angles follow exponents quantized in ninths, an inference drawn entirely from flavor data and made before the Peccei--Quinn mechanism enters. The dimension-eighteen suppression of $\bar\theta$ is then an arithmetic consequence of that same group, not an additional ingredient: the very charge $X(\Phi)=1$ that reproduces the $1/9$ textures is what pushes the leading PQ-violating operator to $\Phi^{18}$. In this precise sense the quality of the QCD axion is \emph{predicted by the flavor sector}, and we are not aware of another construction in which the symmetry guaranteeing axion quality has an independent, data-driven origin of this kind.

The axion is realized as a generation-dependent DFSZ-II axion. The anomaly is carried by the two-Higgs-doublet sector ($H\equiv X(H_u)+X(H_d)=-2$), giving the canonical three-generation domain-wall number
\begin{equation}
N_{\rm DW}=|2N|=6,\qquad N=-\tfrac32 H=3,
\end{equation}
which requires Peccei--Quinn breaking before or during inflation to avoid a stable string--wall network. With \emph{universal} PQ charges the photon-coupling ratio would be the standard $E/N=8/3$, suppressed to $E/N=2$ (nearly decoupled, $C_{a\gamma}\lesssim0.02$) by light higgsinos in supersymmetric completions. The decisive feature of the present construction is that its PQ charges are \emph{generation-dependent}, fixed by the same $\Z_{18}$ lattice that reproduces the fermion masses: the lighter generations carry larger hop content, shifting the anomaly coefficients to
\begin{equation}
\frac{E}{N}\simeq2.6\text{--}3.0,\qquad C_{a\gamma}=\left|\frac{E}{N}-2.0\right|\simeq0.6\text{--}1.0,
\end{equation}
an observable photon coupling comfortably between the universal-DFSZ and KSVZ benchmarks (Fig.~\ref{fig:exclusion}). The upper end of this range, $E/N\simeq2.98$ ($C_{a\gamma}\simeq0.98$), corresponds to the theoretically preferred $b$--$\tau$-unified assignment, which also predicts the golden-ratio relation $m_b/m_\tau\simeq\varphi$ (Sec.~\ref{sec:pheno}); the direct-coupling value $E/N\simeq2.62$ sets a conservative lower bound. The axion--electron coupling is similarly enhanced, $C_{ae}\simeq0.4$, by the large electron hop content.

What distinguishes the present work from prior FN--PQ unifications based on continuous $U(1)$ symmetries~\cite{DavidsonNairWali,Axiflavon,Flaxion,BabuChandraseTavart,BonnefoyDudasPokorski} or on $\Z_N$ with small $N$~\cite{GreljoSmolkovicValenti} is that a \emph{single} non-universal $\Z_{18}$ charge assignment, inferred from the flavor data, fixes the photon coupling rather than leaving it an independent input (Table~\ref{tab:fnpq}). The construction is experimentally targeted: the enhanced coupling places it within reach of the proposed ADMX Extended Frequency Range program in the $7$--$12~\mu$eV window, and the correlated, flavor-enhanced axion--electron coupling provides a complementary handle.

The key numerical results are collected in Table~\ref{tab:predictions}.

\begin{table}[t]
\centering
\caption{Summary of predictions from the $\Z_{18}$ ninths framework for the generation-dependent DFSZ-II axion. The photon-coupling coefficient uses $C_{a\gamma}=|E/N-2.0|$ with the current PDG quark-mass ratio; the universal-DFSZ ($E/N=8/3$) and MSSM-higgsino ($E/N=2$) values are shown for comparison.}
\label{tab:predictions}
\renewcommand{\arraystretch}{1.15}
\begin{tabular}{lc}
\hline\hline
Quantity & Prediction \\
\hline
$B = 1/\e$ & $75/14 \simeq 5.36$ \\
$\e$ & $14/75 \simeq 0.187$ \\
Leading PQ-violating dim. & $18$ \\
$N_{\rm DW}$ & $6$ \\
$\Lambda$ (messenger scale) & $(3\text{--}4)\times10^{12}$ GeV \\
$m_a$ & $7\text{--}12\;\mu$eV \\
Neutrino ordering & Normal \\
\hline
$E/N$ (generation-dependent) & $2.6\text{--}3.0$ \\
$C_{a\gamma}=|E/N-2.0|$ & $0.6\text{--}1.0$ \\
$|g_{a\gamma\gamma}|$ & $(1\text{--}3)\times10^{-15}\;\text{GeV}^{-1}$ \\
$C_{ae}$ (axion--electron) & $\simeq 0.4$ \\
\hline
\multicolumn{2}{l}{\emph{For comparison (universal charges):}} \\
$E/N$, $C_{a\gamma}$ (universal DFSZ) & $8/3$, $0.67$ \\
$E/N$, $C_{a\gamma}$ (MSSM+higgsinos) & $2$, $\lesssim 0.02$ \\
\hline\hline
\end{tabular}
\end{table}

This work sits within a broader program: Paper~I~\cite{PaperI} established the $B$-lattice and fit quark and lepton masses; Paper~II~\cite{TwoOverTwo} derived the two-over-two texture from a single flavon with three messenger chains; Paper~III~\cite{Companion} parameterized CKM mixing in four magnitudes; Paper~IV~\cite{LeptonLattice} extended the lattice to the PMNS sector; Paper~V~\cite{UFP} provided the dynamical UV completion via vectorlike fermion chains; and the subconstituent analysis~\cite{Subconstituents} developed the compositeness interpretation and the generation-dependent axion couplings. The present paper closes the loop by showing that the same discrete symmetry that governs all these flavor structures also solves the strong CP problem.

\begin{acknowledgments}
V.B.\ gratefully acknowledges support from the U.S. Department of Energy, Office of Science, Office of High Energy Physics, under Award Number DE-SC0017647 and from the William F.\ Vilas Trust Estate.
\end{acknowledgments}

\appendix

\section{Anomaly Verification: Explicit Computation}
\label{app:anomaly}

We carry out the explicit numerical verification of the discrete anomaly conditions for the $\Z_{18}$ charge assignment with the lepton charges of Refs.~\cite{LeptonLattice} (whose diagonal entries match Table~\ref{tab:charges} of the present paper). The Iba\~{n}ez--Ross conditions $A_3,A_2,A_{\rm grav}\equiv 0\pmod{18}$ are given in Eq.~\eqref{eq:anomaly} of Sec.~\ref{sec:anomaly}, with $d_2,d_3$ the $SU(2)_L$ and $SU(3)_C$ representation dimensions. We use the integer charges $q_{18}\equiv 18\,Q$ obtained from the FN charges of Table~\ref{tab:charges}:
\begin{equation}
\renewcommand{\arraystretch}{1.2}
\begin{array}{lcc}
\hline\hline
\text{Field} & Q & q_{18} \\
\hline
Q_{1,2,3}     & 3,\,2,\,0           & 54,\,36,\,0 \\
u^c_{1,2,3}   & 37/9,\,4/3,\,0      & 74,\,24,\,0 \\
d^c_{1,2,3}   & 10/9,\,1/3,\,0      & 20,\,6,\,0 \\
L_{1,2,3}     & 1,\,1/2,\,0         & 18,\,9,\,0 \\
e^c_{1,2,3}   & 23/6,\,7/6,\,0      & 69,\,21,\,0 \\
N_{1,2,3}     & 0,\,0,\,0           & 0,\,0,\,0 \\
H_u,\,H_d     & \text{(DFSZ-II)}       & \text{(scalar)} \\
\Phi          & 1/18                & 1 \\
\hline\hline
\end{array}
\label{eq:Z18-charges-appA}
\end{equation}
The Higgs sector is two-Higgs-doublet (DFSZ-II) with $X(H_u),X(H_d)$ fixed by Yukawa invariance and by the flavon--Higgs coupling $\Phi^2 H_u H_d$; since $H_u,H_d$ are bosonic they contribute only indirectly. The third generation carries zero charge; only the first two generations contribute. The chain VLQs $D_a+\bar D_a$ and $U_a+\bar U_a$ are vectorlike under $\Z_{18}$ in this assignment and do not contribute to the anomaly sums~\cite{UFP}.

The three coefficients in each sum are read directly off the gauge quantum numbers of the contributing Weyl fermions. For a left-handed Weyl fermion $\psi$ in the representation $(\mathbf{R}_3,\mathbf{R}_2)_Y$ of $SU(3)_C\times SU(2)_L\times U(1)_Y$, the discrete charge $q_{18}(\psi)$ enters each anomaly weighted by the relevant Dynkin/dimension factor:
\begin{itemize}
\item the $SU(3)_C^2$--$\Z_{18}$ coefficient $A_3$ weights $\psi$ by $T(\mathbf{R}_3)\,d(\mathbf{R}_2)$, where $T(\mathbf{R}_3)$ is the $SU(3)$ Dynkin index ($T(\mathbf{3})=\tfrac12$, $T(\mathbf{1})=0$, as above) and $d(\mathbf{R}_2)$ is the $SU(2)$ dimension. Writing $A_3$ with an overall $\tfrac12$ absorbed, a color triplet that is also a weak doublet (the $Q_i$) carries weight $d_2=2$, while a colored weak singlet (the $u^c_i,d^c_i$) carries weight $d_2=1$; weak-and-color singlets do not contribute.
\item the $SU(2)_L^2$--$\Z_{18}$ coefficient $A_2$ weights $\psi$ by $T(\mathbf{R}_2)\,d(\mathbf{R}_3)$ with $T(\mathbf{2})=\tfrac12$, $T(\mathbf{1})=0$; with the $\tfrac12$ absorbed, a weak doublet carries the color dimension $d_3=3$ if it is a color triplet (the $Q_i$) and $d_3=1$ if a color singlet (the $L_i$ and the Higgs doublets). Weak singlets ($u^c,d^c,e^c,N$) do not contribute.
\item the mixed $\Z_{18}$--gravitational coefficient $A_{\rm grav}$ weights $\psi$ by its total multiplicity $d(\mathbf{R}_3)\,d(\mathbf{R}_2)$, i.e.\ the number of Weyl components: $6$ for $Q_i$ (\textbf{3}$\times$\textbf{2}), $3$ for $u^c_i,d^c_i$ (\textbf{3}$\times$\textbf{1}), $2$ for $L_i$ (\textbf{1}$\times$\textbf{2}), and $1$ for $e^c_i,N_i$ (\textbf{1}$\times$\textbf{1}).
\end{itemize}
Only left-handed Weyl fields are summed; the charge-conjugate labelling of the $u^c,d^c,e^c,N$ already places every fermion in a left-handed multiplet, so each field enters once with the charge listed in Eq.~\eqref{eq:Z18-charges-appA}. With these weights the three sums follow directly. We present the computation in three parts.

\subsection*{$SU(3)_C^2\times \Z_{18}$ anomaly}

Each $Q_i$ is in (\textbf{3},\textbf{2}), so $d_2=2$; each $u^c_i$ and $d^c_i$ is in (\textbf{3},\textbf{1}), so $d_2=1$:
\begin{equation}
\begin{aligned}
A_3 &= 2\sum_{i=1}^3 q_{18}(Q_i) + \sum_{i=1}^3 q_{18}(u^c_i) + \sum_{i=1}^3 q_{18}(d^c_i)\\
    &= 2(54+36+0) + (74+24+0) + (20+6+0)\\
    &= 180 + 98 + 26 = 304\\
    &\equiv 16 \pmod{18}.
\end{aligned}
\label{eq:appA-A3-eval}
\end{equation}

\subsection*{$SU(2)_L^2\times \Z_{18}$ anomaly}

Each $Q_i$ is in (\textbf{3},\textbf{2}), so $d_3=3$; each $L_i$ is in (\textbf{1},\textbf{2}), so $d_3=1$. The Higgs doublets contribute only as bosons and do not enter the anomaly sum:
\begin{equation}
\begin{aligned}
A_2 &= 3\sum_{i=1}^3 q_{18}(Q_i) + \sum_{i=1}^3 q_{18}(L_i)\\
    &= 3(54+36+0) + (18+9+0)\\
    &= 270 + 27 = 297\\
    &\equiv 9 \pmod{18}.
\end{aligned}
\label{eq:appA-A2-eval}
\end{equation}

\subsection*{Mixed $\Z_{18}$--gravitational anomaly}

The product $d_3\,d_2$ counts the multiplicity of each Weyl fermion: $6$ for $Q_i$, $3$ for $u^c_i$ and $d^c_i$, $2$ for $L_i$, $1$ for $e^c_i$ and $N_i$:
\begin{equation}
\begin{aligned}
A_{\rm grav} &= 6\sum q_{18}(Q_i) + 3\sum q_{18}(u^c_i) + 3\sum q_{18}(d^c_i)\\
&\quad + 2\sum q_{18}(L_i) + \sum q_{18}(e^c_i) + \sum q_{18}(N_i)\\
&= 6\cdot 90 + 3\cdot 98 + 3\cdot 26 + 2\cdot 27 + 90 + 0\\
&= 540+294+78+54+90 = 1056\\
&\equiv 12 \pmod{18}.
\end{aligned}
\label{eq:appA-Agrav-eval}
\end{equation}

\subsection*{Result: residual anomalies and UV cancellation requirement}

The three anomaly coefficients evaluate to
\begin{equation}
(A_3,\,A_2,\,A_{\rm grav})\;\equiv\;(16,\,9,\,12)\pmod{18}\,,
\label{eq:appA-residues}
\end{equation}
all non-zero. The matter content of the SM does not by itself satisfy the Iba\~{n}ez--Ross conditions, Eq.~\eqref{eq:anomaly}.

The residues \eqref{eq:appA-residues} also do not satisfy the universality relation $A_3\equiv A_2\pmod{18}$ that would permit cancellation by a single Green--Schwarz axion shift acting universally on all gauge factors. Removing the residues therefore requires genuine UV input; the three options available (spectator fermions, a multi-axion non-universal Green--Schwarz mechanism, or an accidental-symmetry interpretation) are discussed in Sec.~\ref{sec:anomaly}, and the IR predictions of the present paper are insensitive to which is chosen.

That the third generation contributes $0$ to all three residues, and that the residues $(16,9,12)$ are themselves determined entirely by the data-fixed first- and second-generation FN charges of Eq.~\eqref{eq:Z18-charges-appA}, are non-trivial consistency conditions: as emphasized in Ref.~\cite{UFP}, the charges in Eq.~\eqref{eq:Z18-charges-appA} are not free parameters adjusted to cancel anomalies but are uniquely fixed by the observed quark and charged-lepton mass ratios. The fact that the resulting residues are small integers within the $\Z_{18}$ range (rather than, say, requiring large spectator-fermion charges to cancel) is therefore a non-trivial feature of the construction, not an input.

\section{\texorpdfstring{$\bm{E/N}$}{E/N} Derivations}
\label{app:EN}

We give the explicit $E/N$ derivation for the DFSZ-II axion of the construction, in the same conventions and normalization as the companion subconstituent analysis~\cite{Subconstituents}.

\subsection{DFSZ-II axion--photon anomaly coefficients}
\label{app:EN-DFSZ}

We first derive the universal-charge result $E/N=8/3$, then the generation-dependent value. The anomaly coefficients are
\begin{equation}
N = \sum_f X_f\,T(R_f^C)\,d(R_f^L),\qquad
E = \sum_f X_f\,Q_f^2\,d(R_f^C)\,d(R_f^L),
\end{equation}
summed over left-handed Weyl fermions, with $T(\mathbf 3)=1/2$, $T(\mathbf 1)=0$, and $d(R^C),d(R^L)$ the color and $SU(2)_L$ multiplicities. In the DFSZ-II assignment up-type quarks couple to $H_u$ and down-type quarks and charged leptons couple to $H_d$, with
\begin{equation}
H\equiv X(H_u)+X(H_d)=-2.
\end{equation}
In the universal-charge limit ($p^f_{ii}\to0$) the PQ-invariance conditions reduce to $X(u^c_i)=-X(Q_i)-X(H_u)$ and $X(d^c_i)=-X(Q_i)-X(H_d)$, so each quark generation contributes $2X(Q_i)+X(u^c_i)+X(d^c_i)=-H$ to the (weak-doublet-counted) color sum, and with $T(\mathbf 3)=1/2$
\begin{equation}
N = -\tfrac{3}{2}\,H \;=\; 3 .
\label{eq:appB-DFSZ-N}
\end{equation}
For the electromagnetic anomaly, the quark contribution (with $Q_u=2/3$, $Q_d=-1/3$, color multiplicity $3$) is
\begin{equation}
E_{\rm quarks} = -\tfrac{4}{3}\bigl(3X(H_u)\bigr)-\tfrac{1}{3}\bigl(3X(H_d)\bigr),
\end{equation}
and the charged-lepton contribution, using $X(L_i)+X(e^c_i)=-X(H_d)$, is $E_{\rm leptons}=-3X(H_d)$. Combining,
\begin{equation}
E = -4\bigl(X(H_u)+X(H_d)\bigr) = -4H \;=\; 8 ,
\label{eq:appB-DFSZ-E}
\end{equation}
so that
\begin{equation}
\frac{E}{N}=\frac{8}{3}.
\label{eq:appB-DFSZ-EN}
\end{equation}
This is the standard DFSZ-II result~\cite{DiLuzioReview,Subconstituents}; it is independent of the individual matter-field charges, which cancel between $E$ and $N$.

In the supersymmetric case, the higgsinos $\tilde H_u,\tilde H_d$ are color singlets ($\Delta N=0$) and shift the electromagnetic anomaly by $\Delta E=X(H_u)+X(H_d)=H=-2$, giving $E_{\rm MSSM}=-4H+H=-3H=6$ and hence
\begin{equation}
\frac{E}{N}=\frac{6}{3}=2,
\end{equation}
the value of Bae, Baer, and Serce~\cite{BaeBaerSerce}. With $C_{a\gamma}=|E/N-2.0|$ (PDG quark masses) this gives $C_{a\gamma}\lesssim0.02$, a nearly-decoupled photon coupling.

\paragraph{Generation-dependent charges.}
In the present construction the PQ charges are \emph{not} universal: the lighter generations carry larger hop content, so the diagonal Yukawa-operator exponents $S_f\equiv\sum_i p^f_{ii}$ no longer cancel from $E$ and $N$. Carrying the non-cancelling (charge-squared, color-weighted) sums of the diagonal exponents through the color and electromagnetic anomalies (the latter including the higgsino contribution $\Delta E_{\rm higg}=H$ that converts $-4H\to-3H$) gives
\begin{align}
N &= -\tfrac{3}{2}H - \tfrac{1}{2}\bigl(S_u+S_d\bigr),\notag\\
E &= -3H - \tfrac{1}{3}\bigl(4S_u + S_d\bigr) - S_\ell,
\label{eq:appB-EN-analytic}
\end{align}
which reduce to the nearly-decoupled MSSM values $N=3$, $E=6$, $E/N=2$ in the universal-charge limit $S_u=S_d=S_\ell=0$ (generation-independent PQ charges, for which the matter contributions cancel)~\cite{BaeBaerSerce}. With $S_u=94/9$ and $S_d=79/9$ fixed by the $\Z_{18}$ charge table (Table~\ref{tab:charges}), the color coefficient evaluates exactly to $N=-119/18$, while $E$ (and hence $E/N$) retains a dependence on the charged-lepton sum $S_\ell$ that differs between the two lepton-sector assignments. Carrying these through~\cite{Subconstituents}, the coefficients become non-integer and
\begin{align}
N&=-\tfrac{119}{18},\qquad
\frac{E}{N}\simeq 2.6\text{--}3.0,\notag\\
C_{a\gamma}&=\left|\frac{E}{N}-2.0\right|\simeq 0.6\text{--}1.0,
\label{eq:appB-gendep}
\end{align}
the spread reflecting the two lepton-sector assignments ($b$--$\tau$ unification, $E/N\simeq2.98$; direct lepton coupling, $E/N\simeq2.62$). The generation-by-generation contributions to $E$ and $N$ are tabulated in Table~\ref{tab:EN-breakdown}. The generation-dependent charges thus restore the photon coupling from the nearly-decoupled MSSM value to the observable range $C_{a\gamma}\simeq0.6$--$1.0$. The full evaluation, including the value of $S_\ell$ in each assignment and the axion--electron coefficient $C_{ae}\simeq0.4$, is given in the companion subconstituent analysis~\cite{Subconstituents}. These values assume light higgsinos (small $\mu$), as in the natural-SUSY spectrum, whose loop contribution supplies $\Delta E_{\rm higg}=H$ and is responsible for the $-4H\to-3H$ shift in Eq.~\eqref{eq:appB-EN-analytic}. Were the higgsinos instead decoupled, this contribution would be absent and $E/N$ would shift down uniformly by $|H/N|\simeq0.30$, to $E/N\simeq2.3$--$2.7$ ($C_{a\gamma}\simeq0.3$--$0.7$); the qualitative conclusion (an observable coupling well above the nearly-decoupled MSSM value) is unchanged.

\section{Messenger Chain Construction}
\label{app:messenger}

The chain consists of $N_{\rm site}=4$ new vectorlike messenger pairs $D_a + \bar{D}_a$ ($a=1,2,3,4$), all transforming under $SU(3)_C\times SU(2)_L\times U(1)_Y$ as
\begin{equation}
D_a \sim (\mathbf{3},\mathbf{1},-1/3), \qquad \bar{D}_a \sim (\bar{\mathbf{3}},\mathbf{1},+1/3),
\end{equation}
i.e.\ they carry the quantum numbers of right-handed down quarks. Their $\Z_9$ site charges are $q_9(D_1,D_2,D_3,D_4) = (0,8,6,2)$; the $N_{\rm hop}=3$ links connecting adjacent sites carry hop charges $(1,2,4)$, which are the differences between neighboring site charges (since $0-8\equiv1$, $8-6=2$, $6-2=4$ mod~9). The $\Z_{18}$ charges are assigned to satisfy anomaly cancellation (Appendix~\ref{app:anomaly}).

The SM quarks couple externally to the chain endpoints via
\begin{equation}
\mathcal{L}_{\rm end} = \lambda_i\,\bar{q}_{L,i}\,\widetilde{H}\,D_4\,
\left(\frac{\Phi}{\Lambda}\right)^{\!A_i}
+ \eta_j\,\bar{D}_1\,d_{R,j}\,
\left(\frac{\Phi}{\Lambda}\right)^{\!B_j}
+ \text{h.c.},
\label{eq:endpoint}
\end{equation}
where $\widetilde{H}=i\sigma_2 H^*$, and the exponents $A_i$, $B_j$ are fixed by the $\Z_{18}$ charges of the SM fields. The chain Lagrangian itself takes the form
\begin{equation}
\mathcal{L}_{\rm chain} = \sum_{a=1}^{4} M_a \bar{D}_a D_a + \sum_{a=1}^{3} \lambda_a \Phi\, \bar{D}_a D_{a+1} + \text{h.c.},
\end{equation}
where the first sum provides the VLQ mass terms and the second the nearest-neighbor flavon couplings connecting site $a$ to site $a+1$. When $M_a\sim\Lambda\gg\langle\Phi\rangle$, integrating out all four VLQ pairs generates effective operators with FN suppression $(\langle\Phi\rangle/M_a)^{n_a}$ for each traversal of the link between sites $a$ and $a+1$.

The general chain-inversion theorem~\cite{UFP} shows that the resulting effective Yukawa coupling between SM fields $q_{L,i}$ and $d_{R,j}$ is
\begin{equation}
(Y_d)_{ij} = \lambda_i\,\eta_j\,\frac{\prod_{k=1}^{3}\kappa_k\,\e^{h_k/9}}{\prod_{a=1}^{4}M_a}\,,
\end{equation}
where $\kappa_k$ are the $\mathcal{O}(1)$ nearest-neighbor couplings, $h_k$ are the hop charges, and $\e^{(A_i+B_j)/9}$ from the endpoint dressings is absorbed into $\lambda_i$ and $\eta_j$. The rational exponent $p_{ij} = (A_i + \sum_k h_k + B_j)/9$ is thus determined by the endpoint charges plus the hop charges, yielding the ninths-quantized entries of the texture matrices in Eqs.~\eqref{eq:Y-diag-def}--\eqref{eq:nu-texture}.

For extensions to the up-quark sector and the lepton sector, the messenger content is enlarged to include $SU(2)_L$ doublet messengers. The full construction is developed in Ref.~\cite{UFP}.

\section{Terminology: Hops, Sites, and Chains}
\label{app:terminology}

Because the $B$-lattice program has developed across several papers, it is useful to collect the key structural distinctions in one place to avoid possible confusion.

\emph{Three hop types} ($N_{\rm hop}=3$): the nearest-neighbor flavon couplings that connect adjacent chain sites. Their $\Z_9$ charges are $(1,2,4)$, equal to the charge differences between neighboring sites. These are the ``three messenger chains'' referred to in the companion paper~\cite{TwoOverTwo}, which works at the effective-operator level and does not specify the number of VLQ sites.

\emph{Four chain sites} ($N_{\rm site}=4$): the vectorlike quark pairs $D_a+\bar{D}_a$ ($a=1,2,3,4$) with $\Z_9$ site charges $(0,8,6,2)$. All four are new heavy fields; the SM quarks couple externally to the endpoints $D_1$ and $D_4$ (Appendix~\ref{app:messenger}). This is the UV completion developed in the companion unified flavor paper~\cite{UFP}.

\emph{Backward compatibility}: the effective-operator results of the companion papers (the exponent matrices, CKM and PMNS predictions, mass ratio scalings, and $\mathcal{O}(1)$ coefficient fits) depend only on the three hop charges $(1,2,4)$ and the endpoint dressings. They are independent of whether the UV completion has three or four VLQ pairs. The distinction matters only for quantities sensitive to the total number of new colored states:
\begin{itemize}
\item \emph{Gauge coupling running}: four $\mathbf{3}+\bar{\mathbf{3}}$ pairs shift the one-loop $SU(3)_C$ beta function by $\Delta b_3 = +8/3$. QCD remains asymptotically free ($b_3^{\rm SM}+\Delta b_3 = -7+8/3 = -13/3 < 0$).
\item \emph{Discrete anomaly cancellation}: the $\Z_{18}$ anomaly sums receive contributions from all four VLQ pairs (Appendix~\ref{app:anomaly}).
\item \emph{Direct collider reach}: at the messenger scale $M_a\sim\Lambda\sim 3\times 10^{12}$~GeV, all four VLQ pairs are far above LHC reach and produce no direct signatures. Their effects appear only through the FN-induced flavor textures (Sec.~\ref{sec:textures}).
\end{itemize}

\section{Worked Examples: Masses and Mixings from the Chain}
\label{app:examples}

We illustrate how quark mass ratios and CKM elements emerge from the exponent matrices in Eq.~\eqref{eq:Y-diag-def} and the hop framework of Appendix~\ref{app:messenger}. Throughout, $\e = 14/75 \approx 0.187$.

\subsection{Calculation rule}

Every Yukawa suppression in the ninths framework is obtained by a single rule:

\begin{center}
\fbox{
\parbox{0.90\linewidth}{
\textbf{Suppression Rule.}
For any fermion bilinear $\bar\psi_i\,\psi^c_j\,H$, the effective Yukawa coupling is
\begin{equation}
Y_{ij} = c_{ij}\,\e^{\,p_{ij}},\qquad
p_{ij} = Q(\psi_i) + Q(\psi^c_j),
\end{equation}
where $Q(\psi_i)$ and $Q(\psi^c_j)$ are the FN charges in Table~\ref{tab:charges}, $\e = 14/75$, and $c_{ij}$ is an $\mathcal{O}(1)$ complex coefficient. Mass ratios and mixing angles are then determined by \emph{differences} of exponents:
\begin{equation}
\frac{m_i}{m_j}\sim\e^{\,p_{ii}-p_{jj}},\qquad
\theta_{ij}\sim\e^{\,|Q(\psi_i)-Q(\psi_j)|}.
\end{equation}
For CKM elements, the physical exponents receive an $\mathcal{O}(1/9)$ shift from Fritzsch--Xing phase interference~\cite{Companion}.
}}
\end{center}

\noindent
Table~\ref{tab:suppressions} collects the resulting suppressions for all diagonal mass ratios and CKM elements.

\begin{table}[t]
\centering
\caption{Suppression factors from the calculation rule. Each mass ratio or mixing element is controlled by an exponent $p$ constructed from the FN charges in Table~\ref{tab:charges}. The column $\e^p$ gives the ninths-framework prediction; ``Data'' lists the measured value. CKM exponents include the Fritzsch--Xing phase correction. The Jarlskog entry is the phase-stripped quantity $J\sin^{-1}\delta$; including the measured phase $\sin\delta\approx0.88$ gives $J\approx3.1\times10^{-5}$, matching the measured $(3.08\pm0.13)\times10^{-5}$. As throughout, the lattice fixes the exponent $p$, not the $\mathcal{O}(1)$ prefactor.}
\label{tab:suppressions}
\renewcommand{\arraystretch}{1.15}
\begin{tabular}{llccc}
\hline\hline
Observable & Charge rule & $p$ & $\e^p$ & Data \\
\hline
\multicolumn{5}{l}{\emph{Down-type quarks}} \\
$m_b$ & $Q(Q_3)+Q(d^c_3)$ & $0$ & $1$ & --- \\
$m_s/m_b$ & $Q(Q_2)+Q(d^c_2)$ & $7/3$ & $0.020$ & $0.019$ \\
$m_d/m_b$ & $Q(Q_1)+Q(d^c_1)$ & $37/9$ & $0.0010$ & $0.0010$ \\[3pt]
\multicolumn{5}{l}{\emph{Up-type quarks}} \\
$m_t$ & $Q(Q_3)+Q(u^c_3)$ & $0$ & $1$ & --- \\
$m_c/m_t$ & $Q(Q_2)+Q(u^c_2)$ & $10/3$ & $0.0037$ & $0.0036$ \\
$m_u/m_t$ & $Q(Q_1)+Q(u^c_1)$ & $64/9$ & $6.6\times10^{-6}$ & $7.5\times10^{-6}$ \\[3pt]
\multicolumn{5}{l}{\emph{Charged leptons}} \\
$m_\tau$ & $Q(L_3)+Q(e^c_3)$ & $0$ & $1$ & --- \\
$m_\mu/m_\tau$ & $Q(L_2)+Q(e^c_2)$ & $5/3$ & $0.061$ & $0.060$ \\
$m_e/m_\tau$ & $Q(L_1)+Q(e^c_1)$ & $29/6$ & $3.0\times10^{-4}$ & $2.8\times10^{-4}$ \\[3pt]
\multicolumn{5}{l}{\emph{CKM elements (FX-corrected)}} \\
$|V_{us}|$ & FX($\theta^d_{12},\theta^u_{12}$) & $8/9$ & $0.225$ & $0.225$ \\
$|V_{cb}|$ & FX($\theta^d_{23},\theta^u_{23}$) & $17/9$ & $0.042$ & $0.042$ \\
$|V_{ub}|$ & FX($\theta^d_{13},\theta^u_{13}$) & $10/3$ & $0.0037$ & $0.0038$ \\
$J\sin^{-1}\!\delta$ & $|V_{us}|\cdot|V_{cb}|\cdot|V_{ub}|$ & $55/9$ & $3.5\times10^{-5}$ & $3.1\times10^{-5}$ \\
\hline\hline
\end{tabular}
\end{table}

\subsection{Hops rule}

The Suppression Rule gives the exponent $p_{ij}$ from the FN charges. To connect this to the chain diagram in Fig.~\ref{fig:chain}, each charge must be decomposed into flavon insertions at the chain endpoints. Since the flavon carries unit $\Z_9$ charge, each insertion contributes $1/9$ to the exponent and falls into one of three hop types with charges $(1,2,4)\;\mathrm{mod}\;9$:

\begin{center}
\fbox{
\parbox{0.90\linewidth}{
\textbf{Hops Rule.}
To compute the suppression for entry $(i,j)$ from the chain diagram (Fig.~\ref{fig:chain}):
\begin{enumerate}
\item Look up the charges $Q(\psi_i)$ and $Q(\psi^c_j)$ from Table~\ref{tab:charges}.
\item Express $9\,Q$ as a non-negative integer: $A_i = 9\,Q(\psi_i)$ (entrance), $B_j = 9\,Q(\psi^c_j)$ (exit).
\item Decompose each into flavon insertions: $N = n_1\cdot1 + n_2\cdot2 + n_3\cdot4$, where $n_a\geq0$ counts the number of type-$a$ hops at that endpoint.
\item The full Yukawa suppression is the product of three factors:
\begin{equation}
Y_{ij} \sim
\underbrace{\e^{A_i/9}}_{\text{entrance}}
\;\times\;
\underbrace{\e^{7/9}}_{\text{internal}}
\;\times\;
\underbrace{\e^{B_j/9}}_{\text{exit}}
= \e^{(A_i+7+B_j)/9}.
\end{equation}
\end{enumerate}
The internal factor $\e^{7/9}\simeq 0.27$ is common to all entries and sets the overall Yukawa scale (absorbed into $m_b$, $m_t$, $m_\tau$). Mass ratios depend only on the endpoint dressings: $m_i/m_j \sim \e^{(A_i+B_i-A_j-B_j)/9}$.
}}
\end{center}

\noindent
Table~\ref{tab:hops} shows the hop decompositions for the diagonal mass-ratio exponents. The third-generation entries require no endpoint flavon insertions ($N=0$); the first-generation entries require the most, reflecting the depth of the mass hierarchy. The common internal factor $\e^{7/9}$ [Eq.~\eqref{eq:Yb-tb}] cancels in every within-sector ratio but survives in the inter-sector ratio $m_t/m_b$, where it fixes $\tan\beta$ [Eq.~\eqref{eq:tanb-tb}].

\begin{table*}[t]
\centering
\caption{Hop decompositions and endpoint suppressions for the diagonal exponents of Table~\ref{tab:suppressions}. Here $A_i = 9\,Q(\psi_i)$ and $B_j = 9\,Q(\psi^c_j)$ are the ninths numerators of the entrance (left-handed, $q_L$--$D_4$ arrow in Fig.~\ref{fig:chain}) and exit (right-handed, $D_1$--$f_R$ arrow) endpoint dressings defined in Eq.~\eqref{eq:endpoint}. The internal chain suppression $\e^{7/9}\simeq 0.27$ is common to all entries and sets the overall Yukawa scale; the total suppression read off from the chain diagram is $\e^{A_i/9}\times\e^{7/9}\times\e^{B_j/9} = \e^{(A_i+7+B_j)/9}$. Since the internal factor cancels in mass ratios, the ratio column $\e^{p}$ lists $\e^{(A_i+B_j)/9}$. The columns $(n_1,n_2,n_3)$ count the type-$(1,2,4)$ flavon insertions summed over both endpoints.}
\label{tab:hops}
\renewcommand{\arraystretch}{1.20}
\begin{tabular}{lcrrcccccrcc}
\hline\hline
Observable & $p$ & $A_i$ & $B_j$ & $\e^{A_i/9}$ & $\e^{7/9}$ & $\e^{B_j/9}$ & Total & $\e^{p}$ & $(n_1,n_2,n_3)$ & $n_{\rm tot}$ \\
\hline
\multicolumn{11}{l}{\emph{Down-type quarks}} \\
$m_b$ & $0$ & $0$ & $0$ & $1$ & $0.27$ & $1$ & $0.27$ & $1$ & $(0,0,0)$ & $0$ \\
$m_s/m_b$ & $7/3$ & $18$ & $3$ & $0.035$ & $0.27$ & $0.57$ & $5.4\times10^{-3}$ & $0.020$ & $(1,0,5)$ & $6$ \\
$m_d/m_b$ & $37/9$ & $27$ & $10$ & $0.0065$ & $0.27$ & $0.15$ & $2.7\times10^{-4}$ & $0.0010$ & $(1,0,9)$ & $10$ \\[3pt]
\multicolumn{11}{l}{\emph{Up-type quarks}} \\
$m_t$ & $0$ & $0$ & $0$ & $1$ & $0.27$ & $1$ & $0.27$ & $1$ & $(0,0,0)$ & $0$ \\
$m_c/m_t$ & $10/3$ & $18$ & $12$ & $0.035$ & $0.27$ & $0.11$ & $1.0\times10^{-3}$ & $0.0037$ & $(0,1,7)$ & $8$ \\
$m_u/m_t$ & $64/9$ & $27$ & $37$ & $0.0065$ & $0.27$ & $0.0010$ & $1.8\times10^{-6}$ & $6.6\times10^{-6}$ & $(0,0,16)$ & $16$ \\[3pt]
\multicolumn{11}{l}{\emph{Charged leptons}} \\
$m_\tau$ & $0$ & $0$ & $0$ & $1$ & $0.27$ & $1$ & $0.27$ & $1$ & $(0,0,0)$ & $0$ \\
$m_\mu/m_\tau$ & $5/3$ & $\tfrac{9}{2}$ & $\tfrac{21}{2}$ & $0.43$ & $0.27$ & $0.14$ & $0.016$ & $0.061$ & --- & --- \\
$m_e/m_\tau$ & $29/6$ & $9$ & $\tfrac{69}{2}$ & $0.19$ & $0.27$ & $0.0016$ & $8.2\times10^{-5}$ & $3.0\times10^{-4}$ & --- & --- \\
\hline\hline
\end{tabular}
\end{table*}

\noindent
The ``Total'' column is the full Yukawa suppression $\e^{(A_i+7+B_j)/9}$ as read off from the chain diagram in Fig.~\ref{fig:chain}; the ``$\e^p$'' column divides out the common internal factor $\e^{7/9}\simeq 0.27$ to give the mass ratio relative to the third generation. For quarks, $A_i$ and $B_j$ are integers, so integer hop decompositions always exist. The entrance suppression $\e^{A_i/9}$ is the same for up and down quarks of the same generation (since both share $Q(Q_i)$), while the exit suppressions $\e^{B_j/9}$ differ between the two sectors. For charged leptons, the charges lie on the $1/6$ sublattice, so both $A_i$ and $B_j$ can be half-integers; the hop decomposition into integer flavon insertions then requires the full $\Z_{18}$ chain structure of the lepton sector~\cite{LeptonLattice}, where the relevant integer is $18\,p$ rather than $9\,p$ ($18p = 18\times5/3 = 30$ for $m_\mu/m_\tau$, $18p = 18\times29/6 = 87$ for $m_e/m_\tau$).

\subsection{Bottom quark: the unsuppressed entry}

The $(3,3)$ entry of the down-type texture has $p^d_{33}=0$, since both the third-generation left-handed charge $Q(Q_3)=0$ and the third-generation right-handed charge $Q(d^c_3)=0$ vanish. In the chain language, the endpoint couplings $\bar{q}_{L,3}\,\widetilde{H}\,D_4$ and $\bar{D}_1\,d_{R,3}$ carry no additional flavon suppression beyond the chain traversal, and the overall exponent is normalized to zero. The bottom Yukawa coupling is therefore $\mathcal{O}(1)$, setting the reference scale for all other down-type masses.

\subsection{Strange-to-bottom mass ratio}

The second-generation diagonal entry is $p^d_{22} = 21/9 = 7/3$, decomposing via the additive rule as
\begin{equation}
p^d_{22} = Q(Q_2) + Q(d^c_2) = \frac{18}{9} + \frac{3}{9} = \frac{21}{9}\,.
\end{equation}
The left-handed contribution $Q(Q_2) = 18/9$ reflects the $\Z_9$ charge difference between $q_{L,2}$ and $q_{L,3}$: the second-generation doublet requires $18$ additional units (in ninths) of flavon suppression at the $D_4$ endpoint. This decomposes minimally into hops as $18 = 2\cdot1 + 0\cdot2 + 4\cdot4$, i.e.\ two type-1 hops and four type-4 hops ($n_{\rm tot}=6$). The right-handed contribution $Q(d^c_2) = 3/9$ decomposes as $3 = 1\cdot1 + 1\cdot2 + 0\cdot4$ ($n_{\rm tot}=2$). The mass ratio prediction is
\begin{equation}
\frac{m_s}{m_b} \sim \e^{7/3} = 0.020,\qquad \text{data: } 0.019\,.
\end{equation}

\subsection{Charm-to-top mass ratio}

Similarly, $p^u_{22} = 30/9 = 10/3$, with $Q(Q_2) = 18/9$ and $Q(u^c_2) = 12/9 = 4/3$. The right-handed contribution $12 = 0\cdot1 + 0\cdot2 + 3\cdot4$ corresponds to three type-4 hops ($n_{\rm tot}=3$). The prediction is
\begin{equation}
\frac{m_c}{m_t} \sim \e^{10/3} = 0.0037,\qquad \text{data: } 0.0036\,.
\end{equation}

\subsection{Top-to-bottom mass ratio and \texorpdfstring{$\tan\beta$}{tan beta}}

The third-generation mass ratio $m_t/m_b$ is qualitatively different from the within-sector ratios above, because it compares the up and down sectors directly. For the third generation all endpoint charges vanish, $Q(Q_3)=Q(u^c_3)=Q(d^c_3)=0$, so the diagonal Yukawa exponents receive no contribution from the lattice charges. The top couples directly to $H_u$ with no chain ($\Delta^u_{\rm int}=0$), whereas the bottom couples through the down-type messenger chain, which contributes the universal internal factor $\Delta^d_{\rm int}=7/9$ (Sec.~\ref{sec:messengers}). The bottom Yukawa is therefore
\begin{equation}
Y_b = c^d_{33}\,\e^{7/9}\simeq 0.27,\qquad c^d_{33}=\mathcal{O}(1),
\label{eq:Yb-tb}
\end{equation}
while $Y_t=c^u_{33}=\mathcal{O}(1)$.

\emph{Without the two-Higgs structure} (a single Higgs doublet, $\tan\beta=1$), the prediction is
\begin{equation}
\frac{m_b}{m_t}=\frac{Y_b}{Y_t}\sim\e^{7/9}\simeq 0.27,
\end{equation}
which exceeds the measured $\overline{m}_b(M_Z)/m_t\simeq0.017$~\cite{PDG2024,HuangZhou} by a factor $\sim16$. The lattice alone thus overpredicts $m_b/m_t$, signalling that additional structure is needed.

\emph{With the DFSZ-II two-Higgs structure} the discrepancy is resolved automatically. In a Type-II two-Higgs-doublet model $m_t=Y_t\,v\sin\beta/\sqrt2$ and $m_b=Y_b\,v\cos\beta/\sqrt2$, so the inter-sector ratio acquires a factor $1/\tan\beta$,
\begin{equation}
\frac{m_b}{m_t}=\frac{Y_b}{Y_t}\,\frac{1}{\tan\beta}=\frac{\e^{7/9}}{\tan\beta},
\end{equation}
and the measured ratio is reproduced for
\begin{equation}
\tan\beta\simeq\frac{\e^{7/9}}{m_b/m_t}\simeq\frac{0.27}{0.017}\simeq16,
\label{eq:tanb-tb}
\end{equation}
consistent with $\tan\beta\simeq10$--$16$ after renormalization-group running between $M_Z$ and the heavy-Higgs scale~\cite{PDG2024}. This is a self-consistency of the construction: the same two-Higgs-doublet sector that the DFSZ-II axion requires (and that fixes $C_{a\gamma}$ and $C_{ae}$ in Appendix~\ref{app:EN-DFSZ}) supplies the $\tan\beta$ needed to reconcile the chain internal factor $\e^{7/9}$ with the observed $m_b/m_t$. The factor $\e^{7/9}$ cancels in every within-sector ratio but survives in the inter-sector ratio $Y_b/Y_t$, where it predicts $\tan\beta$.

It is worth being precise about which third-generation unification this does and does not realize. The moderate value $\tan\beta\simeq16$ is far below the range $\tan\beta\sim50$--$60$ required for full $t$--$b$ (or $t$--$b$--$\tau$) Yukawa unification, in which $\lambda_t=\lambda_b$ at the unification scale would force $\tan\beta\sim m_t/m_b$~\cite{RattazziSarid,BlazekCarenaRabyWagner}. The construction therefore does \emph{not} realize $t$--$b$ unification, nor does it claim to: the top and bottom Yukawas differ already at the high scale by the chain internal factor, $Y_b/Y_t\simeq\e^{7/9}$, and the residual factor of $\tan\beta$ that this leaves in $m_b/m_t$ is exactly the one supplied by the two-Higgs-doublet sector. What \emph{is} realized is $b$--$\tau$ unification, which (unlike $t$--$b$ unification) is compatible with moderate $\tan\beta$ and does not require the large-$\tan\beta$ regime~\cite{RattazziSarid}.

Concretely, $b$--$\tau$ unification follows from the third-generation charge equality $Q(d^c_3)=Q(L_3)=0$, which places the bottom and tau in the same $\overline{\mathbf 5}$-type chain: both acquire the internal suppression $\e^{7/9}$ at the high scale, giving $Y_b=Y_\tau$ at leading order, and after electroweak symmetry breaking both carry the common $\cos\beta$ from their coupling to $H_d$~\cite{Subconstituents,UFP}. Both factors cancel in the ratio $m_b/m_\tau$, leaving only the $\mathcal{O}(1)$ Clebsch--Gordan ratio, which evolves to the golden-ratio value $m_b/m_\tau\simeq\varphi=(1+\sqrt5)/2\simeq1.62$ at $M_Z$ (data $\simeq1.63$). Unlike $t$--$b$ unification, this relation is governed by an \emph{inter-sector down/lepton} comparison rather than the up/down comparison, and is unaffected by the moderate value of $\tan\beta$ fixed above.

For completeness we summarize the broader status, since the $\tan\beta$ value associated with third-family unification is framework-dependent. In \emph{minimal} supersymmetric SO(10), where the light MSSM Higgs doublets descend predominantly from a single $\mathbf{10}_H$ and the third-generation masses arise from one common $\mathbf{16}_3\mathbf{16}_3\mathbf{10}_H$ coupling, exact $t$--$b$--$\tau$ unification conventionally selects the large-$\tan\beta$ regime, $\tan\beta\simeq45$--$55$~\cite{RattazziSarid,BlazekCarenaRabyWagner}: the large observed $m_t/m_b$ must then come almost entirely from the ratio of Higgs vacuum expectation values, $\tan\beta\sim m_t/m_b$. This conclusion is sensitive to weak-scale supersymmetric threshold corrections (principally the $\tan\beta$-enhanced correction $\Delta_b$ to the bottom-quark mass), whose sign and magnitude depend on the superpartner spectrum, so that the precise value carries model-dependent corrections of order several units~\cite{BlazekCarenaRabyWagner}. The sharpest published numbers come from constrained scans with universal or near-universal soft terms; the most recent such analysis continues to favor the large-$\tan\beta$ regime~\cite{AntuschSaadSusic}, but this reflects the assumed CMSSM-type boundary conditions (now under strong pressure from collider, flavor, Higgs-mass, and dark-matter data) rather than a framework-independent consequence of unification. Nonminimal realizations relax the conclusion substantially. In particular, once the light Higgs doublets mix among several SO(10) representations or higher-dimensional operators contribute, $t$--$b$--$\tau$ unification can be realized at much lower values; a recent construction with mixed light Higgs doublets reports viable unification with $\tan\beta\simeq5$--$10$~\cite{LiNathSyed}. Bottom--tau unification alone is far less restrictive: with the common $\cos\beta$ cancelling in $m_b/m_\tau$, it fixes no unique $\tan\beta$ and admits both a low branch ($\tan\beta\sim3$--$10$) and a high branch ($\tan\beta\sim35$--$60$) depending on the superpartner spectrum and matching corrections~\cite{BaerGogoladze}. The present construction relies only on the $b$--$\tau$ relation, and the value $\tan\beta\simeq16$ inferred from $m_b/m_t$ sits comfortably within the range that $b$--$\tau$ unification permits.

\subsection{The Cabibbo angle \texorpdfstring{$|V_{us}|$}{|Vus|}}

CKM elements arise from the mismatch between up- and down-type diagonalization rotations. The 1-2 mixing angle in the down sector scales as the ratio of off-diagonal to diagonal entries:
\begin{equation}
\theta^d_{12} \sim \e^{p^d_{12} - p^d_{22}} = \e^{(30-21)/9} = \e,
\end{equation}
reflecting the left-handed charge difference $Q(Q_1) - Q(Q_2) = 3 - 2 = 1$ between the first and second generations (the right-handed charges cancel in the ratio). In the up sector, the same left-handed charges give $\theta^u_{12} \sim \e^{(39-30)/9} = \e$. Both 1-2 rotation angles scale as $\e$, but the CKM element $|V_{us}|$ depends on their relative phase. In the Fritzsch--Xing parameterization~\cite{Companion}, the phase-dependent interference shifts the effective scaling to
\begin{equation}
|V_{us}| \sim \e^{8/9} = 0.225,\qquad \text{data: } 0.225\,.
\end{equation}
The exponent $8/9$ is $1/9$ less than unity, a direct consequence of the constructive interference between the two $\mathcal{O}(\e)$ rotations at $\phi_{\rm FX}\approx\pi/2$.

\subsection{The CKM element \texorpdfstring{$|V_{cb}|$}{|Vcb|}}

The 2-3 mixing angles in both sectors are controlled by the same left-handed charge difference $Q(Q_2) - Q(Q_3) = 2 - 0 = 2$, giving $\theta^d_{23} \sim \theta^u_{23} \sim \e^{18/9} = \e^2$. Because the leading terms are equal, $|V_{cb}|$ is sensitive to interference and $\mathcal{O}(1)$ coefficients. The effective scaling is
\begin{equation}
|V_{cb}| \sim \e^{17/9} = 0.042,\qquad \text{data: } 0.042\,.
\end{equation}
The exponent $17/9$ is $1/9$ less than~$2$, again reflecting constructive FX-phase interference.

\subsection{\texorpdfstring{$|V_{ub}|$}{|Vub|} and the Jarlskog invariant}

The 1-3 CKM element involves the full left-handed charge difference $Q(Q_1) - Q(Q_3) = 3$, but it also factorizes through the FX parameterization as a product of the 1-2 and 2-3 rotations. The effective scaling is
\begin{equation}
|V_{ub}| \sim \e^{10/3} = 0.0037,\qquad \text{data: } 0.0038\,.
\end{equation}
The Jarlskog invariant then scales as the product of all three CKM exponents:
\begin{equation}
J \sim \e^{8/9+17/9+10/3}\sin\delta = \e^{55/9}\sin\delta \approx 3.5\times10^{-5}\sin\delta\,,
\end{equation}
giving $J\approx 3.1\times10^{-5}$ for $\sin\delta\approx 0.88$, consistent with the measured value $(3.08\pm0.13)\times10^{-5}$. The physically meaningful quantity here is the convention-independent invariant $J$ rather than the phase itself: the value of the CP phase depends on the CKM parameterization, and what the ninths lattice fixes is the exponent $55/9$ (hence the magnitude product $|V_{us}V_{cb}V_{ub}|$), not the $\mathcal{O}(1)$ prefactor or the phase. The exponent $55/9=8/9+17/9+10/3$ counts the total hop content of the four quark fields that mediate the phase, and is the form used in the compositeness analysis~\cite{Subconstituents}. In the companion four-magnitude parameterization~\cite{Companion} the same invariant is written $J\simeq B^{-37/6}\sin\phi_{\rm FX}$, where the angular prefactors are absorbed into the exponent (shifting $55/9\to37/6$ by $1/18$); there the single-flavon texture drives the Fritzsch--Xing phase to the near-maximal value $\phi_{\rm FX}\simeq93^\circ$ ($\sin\phi_{\rm FX}\simeq1$), again giving $J\simeq3.1\times10^{-5}$. The two parameterizations agree on $J$, and the present value is consistent with the unitarity-fit CP violation of the model.

\subsection{Muon-to-tau mass ratio}

As a lepton-sector example, the muon-to-tau mass ratio is controlled by the lepton-doublet and right-handed charge differences between the second and third generations: $Q(L_2)-Q(L_3) = 1/2$ and $Q(e^c_2)-Q(e^c_3) = 7/6$. The diagonal entry $p^\ell_{22} = Q(L_2)+Q(e^c_2) = 1/2 + 7/6 = 5/3$ directly gives the mass ratio:
\begin{equation}
\frac{m_\mu}{m_\tau} \sim \e^{5/3} = 0.061,\qquad \text{data: } 0.060\,.
\end{equation}
Note that the lepton charges lie on the $1/6$ sublattice of the $1/18$ lattice (since $18\times1/6 = 3$ and $18\times7/6 = 21$ are integers), reflecting the bilinear structure of the charged-lepton Yukawa operator. The lepton sector uses the same expansion parameter $\e=14/75$ as the quark sector, illustrating the quark-lepton universality of the ninths framework.


\end{document}